\documentclass[superscriptaddress,twocolumn,floatfix,nofootinbib,10.5pt,prx]{revtex4}
\usepackage{graphicx,amsfonts,enumerate}
\usepackage{slashed}
\usepackage{amsmath,amssymb,bm,graphicx,dsfont}
\usepackage[latin9]{inputenc}
\usepackage{amsmath}
\usepackage{slashed}
\usepackage{dsfont}
\usepackage{amstext}
\usepackage{amsbsy}
\usepackage{amsthm}
\usepackage{epsfig}
\usepackage{graphicx, float}
\usepackage{bbm}
\usepackage{xcolor}
\usepackage[colorlinks, linkcolor=blue, citecolor=blue]{hyperref}
\usepackage{hypcap}
\usepackage{soul}

\renewcommand{\v}[1]{\mathbf{#1}} 
\newcommand{\uv}[1]{\mathbf{\hat{#1}}} 

\newcommand{\s}[1]{\mathcal{#1}}

\usepackage{dcolumn}
\usepackage{bm}

 \renewcommand{\vec}[1]{{\bf #1}}
\renewcommand{\hat}[1]{{\widehat #1}}

\def\nn{\nonumber\\}

\allowdisplaybreaks

\newcommand{\td}{\mathrm{d}}

\begin{document}
\title{Interpreting Angle Dependent Magnetoresistance in Layered Materials: \\Application to Cuprates}
\author{Seth Musser}
\affiliation{Department of Physics, Massachusetts Institute of
Technology, Cambridge, MA 02139, USA}
\author{Debanjan Chowdhury}
\affiliation{Department of Physics, Cornell University, Ithaca, NY 14853, USA}
 \author{Patrick A. Lee}
\affiliation{Department of Physics, Massachusetts Institute of
Technology, Cambridge, MA 02139, USA}
 \author{T. Senthil}
\affiliation{Department of Physics, Massachusetts Institute of
Technology, Cambridge, MA 02139, USA}
\date{\today}

\preprint{APS/123-QED}

\begin{abstract}
 The evolution of the low temperature electronic structure of the cuprate metals from the overdoped to the underdoped side has recently been addressed through Angle-Dependant Magneto-Resistance  (ADMR) experiments in La$_{1.6-x}$Nd$_{0.4}$Sr$_x$CuO$_4$. The results show a striking difference between hole dopings $p = 0.24$ and $p = 0.21$ which lie on either side of a putative quantum critical point at intermediate $p$. Motivated by this, we here study the theory of ADMR in correlated layered materials, paying special attention to the role of angle dependent quasiparticle weights $Z_\v{k}$. Such a $Z_\v{k}$ is expected to characterize a number of popular models of the cuprate materials, particularly when underdoped. Further, in the limit of weak interlayer hopping the quasiparticle weight will affect the $c$-axis transport measured in ADMR experiments. We show that proper inclusion of the quasiparticle weight does not support an interpretation of the data in terms of a $(\pi, \pi)$ spin density wave ordered state, in agreement with the lack of direct evidence for such order. We show that a simple model of Fermi surface reconfiguring across a van Hove point captures many of the striking differences seen between $p = 0.21$ and $p = 0.24$. 
 We comment on why such a model may be appropriate for interpreting the ADMR data, despite having a large Fermi surface at $p = 0.21$, seemingly in contradiction with other evidence for a small Fermi surafce at that doping level. 

\end{abstract}

\maketitle

\section{Introduction}
The normal metallic states of the cuprate high temperature superconductors have been a continuing challenge to our understanding of these materials. On the overdoped side there is a well developed Fermi surface whose area (per copper-oxygen plane) matches the expectation from the electron density based on Luttinger's theorem \cite{proust_remarkable_2019}. The underdoped side has the well-known, though not well-understood, pseudogap phenomenon where electronic states near the anti-nodal part of the
Brillouin zone get gapped out at some temperature scale $T^*$ \cite{lee2006doping}. In between these two regimes there is the mysterious strange metal where electronic quasiparticles are apparently destroyed \cite{proust_remarkable_2019}.

Clearly it is extremely important to have direct experimental probes of the electronic structure of the pseudogap state in the low temperature limit, and to know experimentally how it evolves into the overdoped metal. An interesting class of experiments are quantum oscillations that probe Landau quantization of charged carriers in a magnetic field. When quantum oscillations are observed they give clear-cut information about the size of the Fermi surface (if it exists). Remarkably, quantum oscillations have been seen in both underdoped and overdoped cuprates and show striking differences \cite{QOreview}. In the underdoped case, in a range of dopings in YBCO,  a small oscillation frequency corresponding to a Fermi surface area of about $2\%$ of the Brillouin zone is seen \cite{sebastian_quantum_2012,QOreview}. On the overdoped side (studied in the material Tl-2201), quantum oscillations are also seen but with a large frequency that corresponds to a Fermi surface with about $65\%$ of the Brillouin zone area \cite{vignolle_quantum_2008}.

It is fascinating to ask how these two very different behaviors evolve into each other as the doping is changed.  Unfortunately, though, it has not been possible to  study this evolution directly through quantum oscillation experiments. One  essential difficulty is that exposing the normal state requires the use of magnetic fields large enough to suppress superconductivity. This becomes progressively harder as optimal doping is approached due to the high superconducting transition temperature. Furthermore the two cuprates mentioned above on either side of optimal doping, YBCO and Tl-2201, cannot be grown at all doping levels that span the full range from underdoped to overdoped. While there exist other members of the cuprate family that do not have such restrictions on doping, they typically are not clean enough to observe quantum oscillations. 

In this paper we are concerned with a class of experiments --- known as angle dependent magneto-resistance (ADMR) ---  that seek to address this evolution {\it without} all the requirements of quantum oscillations. ADMR refers to the change in the out-of-plane resistivity when an external magnetic field is rotated away from the $c$-axis toward the $ab$-plane. Unlike quantum oscillations, ADMR arises as a semi-classical effect of electron trajectories in orbits around the Fermi surface \cite{kartsovnik_angular_1992}. Since full orbits are not necessary to observe ADMR it has much less stringent requirements on sample purity and low temperatures than quantum oscillations. It can thus be used to obtain information about the Fermi surface, and about the scattering rate of the quasiparticles at the Fermi surface in the weak magnetic field limit \cite{m._abdel-jawad_anisotropic_2006, kennett_sensitivity_2007}. However, extracting this information from ADMR requires a detailed theoretical model in this limit. Nevertheless, ADMR has the potential to be a useful tool to study the Fermi surface structure even in situations where quantum oscillations cannot be measured. 

With these motivations a remarkable recent paper reported the evolution of ADMR in La$_{1.6-x}$Nd$_{0.4}$Sr$_x$CuO$_4$ (Nd-LSCO) at two different doping levels, $p = 0.24$ and $p = 0.21$, measured at $25$K in a 45 tesla magnetic field\footnote{In this material $p=x$ to within $0.005$ \cite{fang_fermi_2020}.} \cite{fang_fermi_2020}. These two dopings lie on either side of a critical doping that has been associated with other changes in the electronic structure in previous experiments \cite{proust_remarkable_2019}. A dramatic difference was reported in the ADMR curves. At $p = 0.24$, as the $B$-field is tilted away from the $c$-axis, the resistivity first shows an upturn followed by a downturn at larger tilt angles. On the other hand, at $p = 0.21$ the resistivity at low tilt angles has a downturn. In both cases at large tilt angles the resistivity starts increasing again, for at least some angles in-plane. Thus at small tilt angles the two doping levels behave strikingly differently while at large tilt angles they are somewhat closer to each other. It was suggested in \cite{fang_fermi_2020} that their results reflected the change from the large overdoped Fermi surface to a small Fermi pocket that likely characterizes the underdoped material. In particular, they suggested that these hole pockets were the result of some kind of reconstruction with an order parameter modulated at $(\pi,\pi)$, e.g. a spin density wave (SDW) reconstruction with this ordering wavevector.

Given the importance of these measurements, it is essential to scrutinize them closely and see what we learn about the evolution of the electronic structure across the critical doping (around $p = 0.23$). In this paper we address several theoretical questions that are pertinent to  the interpretation of ADMR in correlated layered materials, and apply the insights gained to the experiments of \cite{fang_fermi_2020}. 

We firstly find theoretically that a $(\pi,\pi)$ SDW modulated Fermi surface reconstruction, with the parameter values quoted in \cite{fang_fermi_2020}, does {\it not} reproduce the ADMR data for the out-of-plane conductivity after including the essential effects of the momentum-dependent quasiparticle residue, $Z_\v{k}$. In this setting, the SDW reconstructs the original two-dimensional Fermi surface and generates a non-trivial $Z_\v{k}$, which measures the overlap between the quasiparticle and bare electron. Starting from the limit of decoupled two-dimensional layers and perturbatively adding a small inter-layer hopping, it is essential to include the effects of the quasiparticle residue as it is the {\it electron} that hops between layers and {\it not} the renormalized quasiparticle. Since the residue along the `backside' of the reconstructed pockets is significantly suppressed, the quasiparticle contribution to the electron current along the $c-$direction from these states will be negligible. Starting from this limit, where the important interaction-induced phenomena are restricted to the two-dimensional layers with a relatively weak hopping along the $c-$direction, should provide a reasonable framework for describing the experiments in Nd-LSCO. An alternative starting point for describing the experiments is to consider a full three-dimensional reconstruction of the Fermi surface. While conceptually similar, the analog of a $Z_\v{k}$ does not appear here explicitly within Boltzmann transport, as we discuss later. However, we find, albeit through a limited sweep through the space of fit-parameters, that this procedure also fails to account for the experimental observations.

Returning to the general issue of Fermi surface reconstruction leading to a pseudogap, there are more exotic scenarios involving hole pockets without any long range magnetic ordering, such as in the fractionalized Fermi liquid (FL$^*$) \cite{senthil2003fractionalized,TS04,QiSachdev,Punk15,chowdhury2016enigma} (a particular example of which is the phenomenological Yang-Rice-Zhang (YRZ) ansatz \cite{yang_phenomenological_2006}) which violate the conventional Luttinger's theorem \cite{Abrikosov:107441,MO00} but satisfy a modified one \cite{TS04,AV04}. While it remains unclear if the experimental results are consistent with such an exotic state, the effects of $Z_\v{k}$ necessarily need to be included in order to describe the out of plane conductivity. Specifically any proposed FL$^*$ state must have a very small $Z_\v k$ on part of its Fermi surface so as to be consistent with the observation of Fermi arcs, as opposed to Fermi pockets, in angle-resolved photoemission experiments. Most of the considerations discussed earlier in the context of the SDW reconstruction, such as the backside of the hole-pocket not contributing to the conductivity, will continue to remain relevant.  

In order to offer a simpler explanation that can partly explain the ADMR data at $p=0.21$, we consider instead a large Fermi surface state. While it is possible, even likely,  that the metallic {\em  ground state} on the underdoped side undergoes Fermi surface reconstruction to a small pocket, the experiments of \cite{fang_fermi_2020} were conducted at temperatures around $25$K, which is comparable to the pseudogap temperature $T^*$ at this doping, $\sim 50-70$K \cite{matt_electron_2015, doiron-leyraud_pseudogap_2017, cyr-choiniere_pseudogap_2018}.  Thus it may not be surprising that some aspects of the ADMR curve on the underdoped side might be understood via the large Fermi surface that reemerges above the pseudogap temperature. In particular, we show that the observed striking change from upturn to downturn at low angles across the critical doping can be understood in terms of the change in the large Fermi surface as it crosses a van-Hove point, without any need for a Fermi surface reconstruction into a small pocket (e.g. due to broken translation symmetry).

\section{Summary of main results}

In our analysis below, we take a number of simplifying limits that are relevant to the Nd-LSCO samples studied in \cite{fang_fermi_2020}. We firstly assume that $t^\perp/\epsilon_F \ll 1$. This allows us to treat the physics of interlayer conductivity as being driven by hopping between only two metallic layers. By doing so we find, within linear response, that the quasiparticle residue will renormalize the interlayer hopping. We secondly consider the particles to be coherent in-plane, that is that $k_Fl \gg 1$, where $l$ is the mean-free path in-plane. This will allow us to neglect a number of irrelevant corrections to our expressions. Finally, we consider the weak magnetic field limit, $\omega_0 \tau \lesssim 1$, so particles complete at most one orbit in-plane. By use of a toy model we find that in this limit ADMR curves are controlled by the average of the interlayer hopping around the Fermi surface; we use this to propose an alternate explanation for the data observed in \cite{fang_fermi_2020}.

\label{sec:summary}
\subsection{Quasiparticle residue and ADMR}

The ADMR measurements in underdoped cuprates raise the very important question of how to properly include the effects of a non-trivial quasiparticle residue, $Z_\v{k}$, for out of plane transport. While $Z_\v{k}$ drops out of transport within Boltzmann theory, which would also be the case for in-plane transport in the present setting, this is not the case for out-of-plane transport as introduced earlier. In the limit of a weak inter-layer hopping (compared to a characteristic intra-layer energy scale), the effects of $Z_\v{k}$ need to be incorporated in a consistent fashion. An extreme example of this is encountered, for example, when a stack of weakly coupled metals undergo a transition into a paramagnetic Mott insulator with fractionalized (`spinon') excitations \cite{zou_dimensional_2016}. For the spinons, $Z_\v{k}=0$, and tunneling between layers vanishes identically. A trivial limit corresponds to a uniform, momentum-independent $Z_\v{k}=Z$, which would lead to an overall rescaling of the interlayer hopping amplitudes. However, it is not a priori obvious how a momentum dependent $Z_\v{k}$ affects the hopping along $c-$direction. 

\begin{figure}
\centering
\includegraphics[width=\columnwidth]{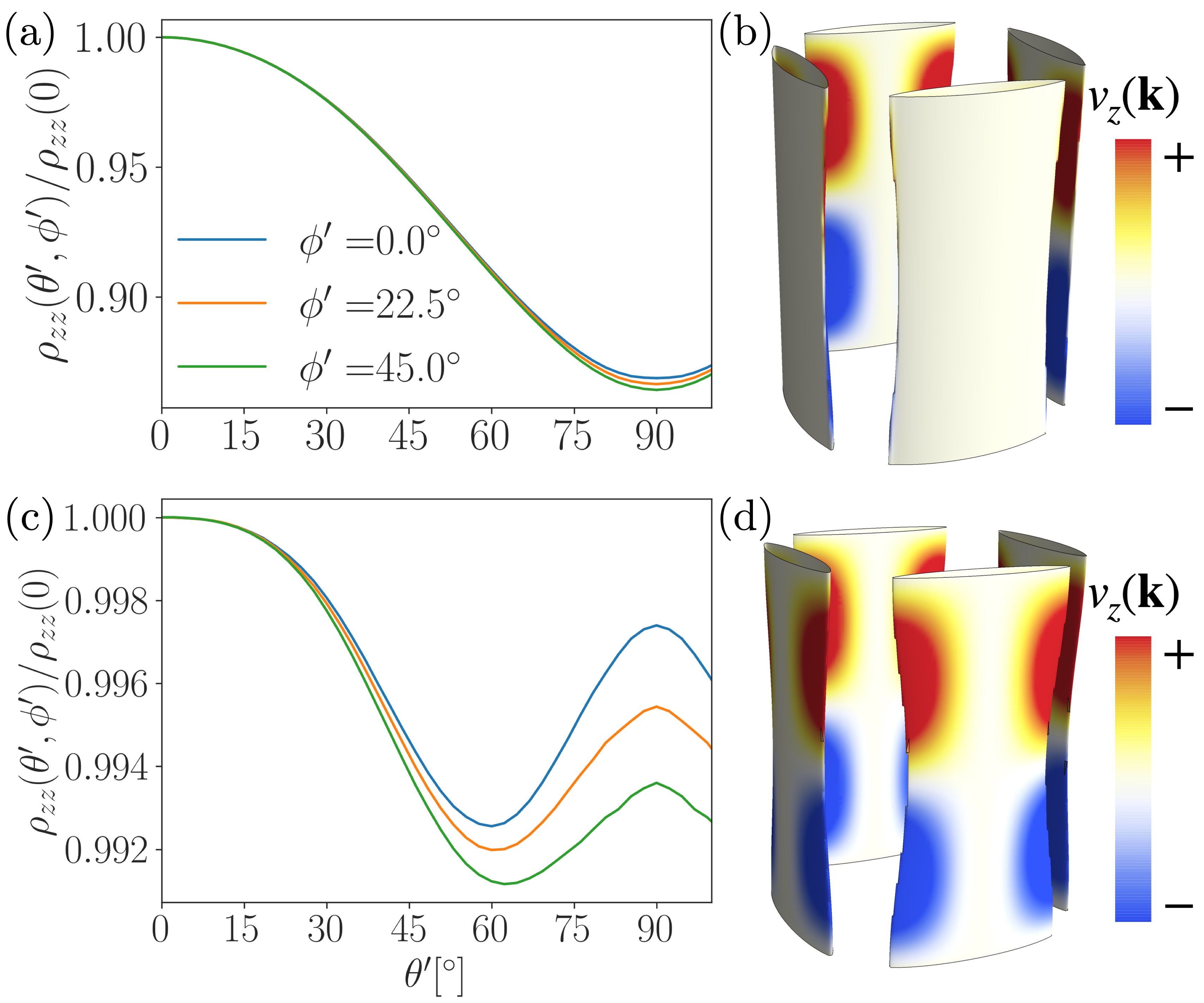}
\caption{ADMR curves for normalized $c-$axis resistivity: (a) including the momentum-dependent $Z_\v{k}$, due to a $(\pi,\pi)$ SDW reconstruction at $p=0.21$. The parameters are otherwise the same as those used in \cite{fang_fermi_2020}; (c) setting $Z_{\v{k}}=1$ in the presence of the same SDW. This reproduces the results in \cite{fang_fermi_2020}.  The electron spectral function and hole pockets: (b) including the non-trivial $Z_{\v{k}}$. The colors represent the value of $v_z(\v{k})$ on the Fermi surface. The backside of the hole pockets are white since $Z_{\v{k}}\rightarrow0$; (d) setting $Z_{\v{k}}=1$. The backside of the hole pockets have an artificial non-zero spectral weight and a finite $v_z(\v{k})$.}
\label{fig:Zk_inclu}
\end{figure}

Microscopically, a momentum-dependent $Z_\v{k}$ can arise due to intrinsic correlation-induced effects \cite{TS08}; its connection to the scattering rate for a momentum independent self-energy has been discussed in Tl-2201 \cite{kokalj_transport_2012}. Alternatively it can also arise from simple band-folding of a Fermi surface (e.g., as in a SDW state). The `coherence-factors' associated with the mean-field diagonalization in the ordered state are precisely responsible for generating the $\v{k}-$dependence of $Z$. Experimentally, there is direct evidence from angle-resolved photoemission measurements \cite{norman_destruction_1998} that underdoped cuprates have a strong $\v{k}$-dependent $Z$. Existing theoretical modeling of ADMR \cite{fang_fermi_2020,yamaji_angle_1989, kartsovnik_angular_1992, moses_comparison_1999} has ignored the effects of angular variation of $Z$ along the Fermi surface without any justification. We fill that conceptual gap in this paper and address the effect of a $Z_{\v{k}}$ within linear response theory in the small inter-layer hopping limit; the technical details are presented in Appendix \ref{sect:Kubo_calc}. We find that within this setup, the interlayer hopping ($t^\perp_{\v{k}}$) is renormalized by $t^\perp_{\v{k}}\rightarrow Z_{\v{k}} t^\perp_{\v{k}}$, which cannot be interpreted as a simple mass renormalization of the carriers along the $c-$direction. As noted earlier, the portions of the Fermi surface with a reduced $Z_{\v{k}}$ (e.g., the backside of the SDW or FL$^*$ hole-pockets) have a significantly reduced contribution to the current along the $c-$direction. As a result, the computations for ADMR lead to dramatically distinct behavior for $c-$axis conductivity with a non-trivial $Z_{\v{k}}$ vs. setting $Z_{\v{k}}=1$, as shown in Figure \ref{fig:Zk_inclu}. Importantly, the proposed $(\pi,\pi)$ SDW \cite{fang_fermi_2020} ceases to explain the data gathered at $p=0.21$ once the effects of $Z_{\v{k}}\neq1$ are incorporated.

It is important to address the evolution of the physical picture introduced above once the strength of the inter-layer hopping is increased, e.g. by varying the ratio of $t^\perp/\epsilon_F$, where $\epsilon_F$ is the in-plane Fermi energy in the decoupled limit. Clearly, as a function of this increasing ratio, there will eventually be a crossover where the full three-dimensional Fermi surface has to be considered from the outset and the problem should not be thought of as quasi two-dimensional. In the three-dimensional limit, $Z_{\v{k}}$ has to drop out of Boltzmann transport in the usual fashion. However, any density-wave ordering in this limit must necessarily reconstruct the full three dimensional Fermi surface, rather than a reconstruction of the in-plane two-dimensional Fermi surface \cite{fang_fermi_2020}. Moreover, it is unclear if a completely uncorrelated ordered SDW along the $c-$direction with a short correlation length would be consistent with the indirect evidence for reasonably coherent semiclassical orbits in ADMR. In  Figure \ref{fig:3d_pipi}, we present our ADMR results for out-of-plane conductivity starting from the full three-dimensional SDW reconstruction of a large Fermi surface, which fails to reproduce the experimental trends observed at $p=0.21$ \cite{fang_fermi_2020}.

\subsection{Physics of ADMR at small tilt angles} 

In light of some of the conceptual challenges that including a non-trivial $Z_\v{k}$ within Boltzmann theory presents, we turn to a much simpler, albeit physically unrealistic, model. The insights from this model will help us  build intuition on how we may interpret various features in ADMR. They will set the stage for our calculations on more realistic models that we  show capture several of the observed features in the experiments. 

Consider a stack of two-dimensional layers with a quadratic in-plane dispersion and hopping along the $z-$direction, $t^\perp_\phi$, between neighboring planes
\begin{equation}
\epsilon(k,\phi,k_z) = \frac{\hbar^2 k^2}{2m^*} - 2t^\perp_\phi \cos(k_za) \label{eqn:toy_model_first},
\end{equation}
where $m^*$ is the effective mass in-plane and $a$ is the interlayer separation. The in-plane Fermi energy in the decoupled limit is denoted $\epsilon_F = \hbar^2 k_F^2/2m^*$, where $k_F$ is the Fermi momentum. This simplified model will allow us to gain intuition about what physics governs the presence of an upturn or downturn at low angles in ADMR and what controls the physics when the field is parallel to the layers. In particular, we will see that when the interlayer hopping averaged over the Fermi surface vanishes, it can support both a sharp downturn at low angles and an upturn at larger angles, similar to the $p=0.21$ data \cite{fang_fermi_2020}.

For $\omega_0\tau\ll 1$, where $\omega_0 = eB/m^*$ and $\tau$ is the quasiparticle lifetime, the appearance of an upturn or downturn in the ADMR curve for $\theta'\ll 1$ (where $\theta'$ is the angle between $B$ and the $c$-axis) is controlled by a competition between the anisotropy of $t^\perp_\phi$ and $(k_Fa)^2/2$.  Increasing the size of the Fermi surface can turn a downturn in the interlayer resistivity to an upturn, while increasing the variation of $t^\perp_\phi$ across the Fermi surface can do the reverse by more effectively averaging the particle's velocity in the $z$-direction, $v_z$, via the in-plane orbits in the limit $\theta'\rightarrow0$. Moreover, a downturn in the ADMR curve at $\theta'\ll 1$ and $\omega_0\tau\ll 1$ evolves as $\omega_0\tau$ is increased, depending on the average $\langle t^\perp_\phi\rangle$ along the Fermi surface; when $\langle t^\perp_\phi \rangle \neq 0$, this initial downturn becomes an upturn with increasing $\omega_0\tau$, while for $\langle t^\perp_\phi \rangle =0$ the downturn survives at large $\omega_0\tau$. 

\begin{figure}
    \centering
    \includegraphics[width=\columnwidth]{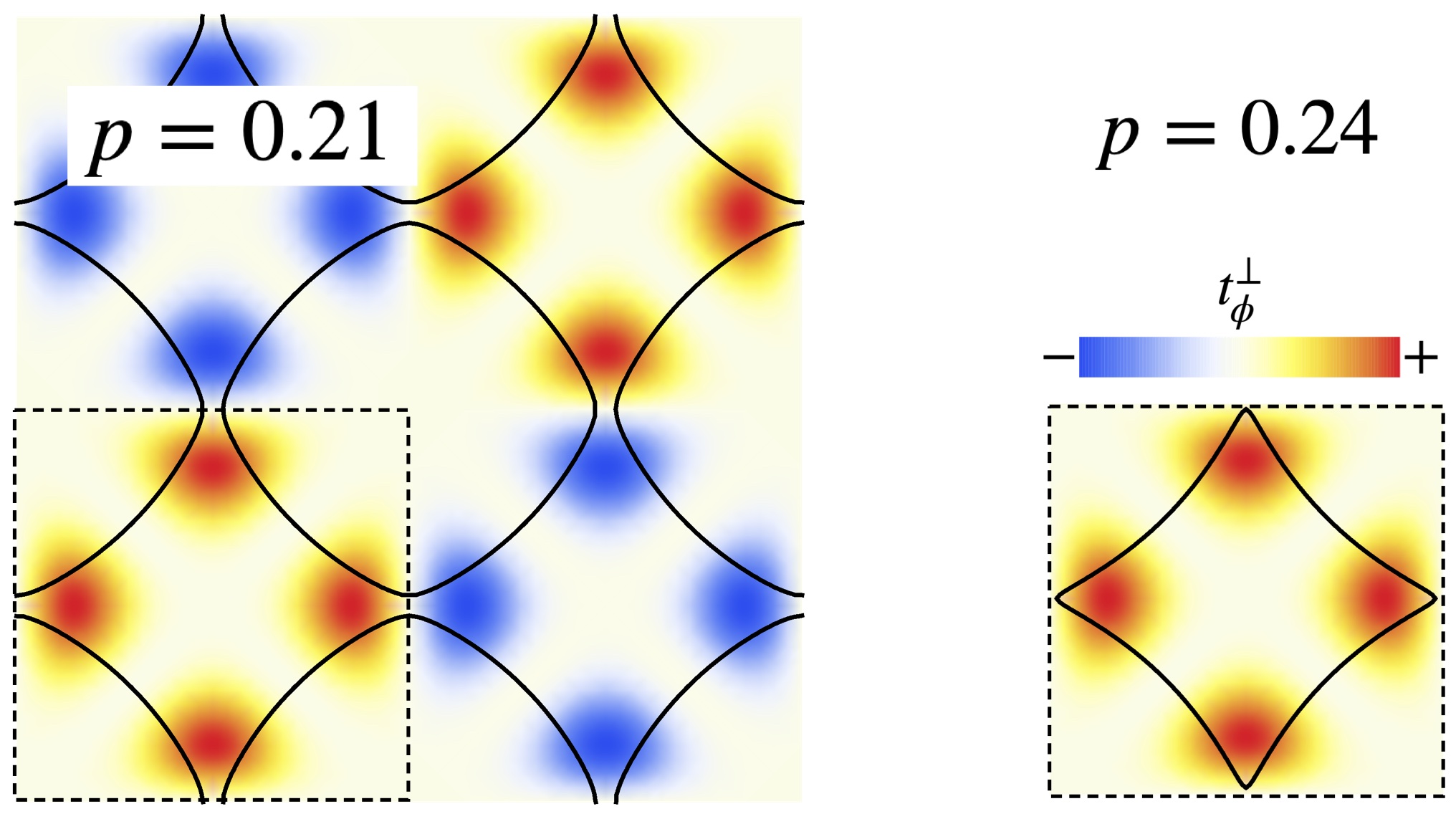}
    \caption{A summary of the change in $t^\perp_\phi$ across the van Hove singularity. The in-plane Fermi surfaces on both the over and underdoped side are plotted with $t^\perp_{\v{k}}$ as in Eq.~\ref{eqn:tperp_k_form} plotted behind. The first Brillouin zone is indicated with a dotted line. On the overdoped side there is no sign change across the Fermi surface, while on the underdoped side the Fermi surface visits areas of negative $t^\perp_\phi$, leading to $\langle t^\perp_\phi\rangle =0$ here.}
    \label{fig:tperp_van_Hove}
\end{figure}

The lessons obtained from analyzing the simplified model reveal the complexity associated with interpreting ADMR results. Interestingly, it also suggests exploring  a potential alternative explanation for the sharp change at low angles in the ADMR curves across $p^*$ \cite{fang_fermi_2020}, without invoking any Fermi surface reconstruction. The van Hove filling in Nd-LSCO is located between the two dopings, $p=0.21$ and $p=0.24$, respectively \cite{matt_electron_2015}. Thus what was a large Fermi surface centered at the $\Gamma$-point on the overdoped side will become a Fermi surface centered at $(\pi,\pi)$ on the underdoped side, assuming no reconstruction occurs at the temperatures where the experiments are carried out. This will substantially change the form of the interlayer hopping across the Fermi surface, as it is given by
\begin{equation}
t_z\cos\left(\frac{k_xa_{\parallel}}{2}\right)\cos\left(\frac{k_ya_{\parallel}}{2}\right)\left[\cos\left(k_xa_{\parallel}\right) - \cos\left(k_ya_{\parallel}\right)\right]^2 \label{eqn:tperp_k_form},
\end{equation}
where $t_z$ is the strength of the interlayer hopping and $a_{\parallel}$ is the in-plane lattice constant. The $\cos(k_xa_{\parallel}/2)\cos(k_ya_{\parallel}/2)$ factor accounts for the offset in the copper oxide planes between layers of the body-centered tetragonal structure of Nd-LSCO \cite{fang_fermi_2020}. For the large overdoped Fermi surface centered at the $\Gamma$-point, this will give rise to a $t^\perp_\phi = t^\perp_{\v{k}_F(\phi)}$ which is positive everywhere and does not average to zero. However, for the underdoped large Fermi surface centered at $(\pi,\pi)$, the $\cos(k_xa_{\parallel}/2)\cos(k_ya_{\parallel}/2)$ factor will ensure that $\langle t^\perp_\phi\rangle = 0$. This is summarized in Figure \ref{fig:tperp_van_Hove}. Based on our results for the simplified model, it is thus plausible that the transition from the upturn at $p=0.24$ to downturn at $p=0.21$ could be driven by a transition from a Fermi surface with $\langle t^\perp_\phi \rangle \neq 0$ to $\langle t^\perp_\phi\rangle =0$. We will show through explicit calculation that this model captures many elements of the data, provided the in-plane Fermi velocity is renormalized to have a slightly larger value at the antinodes.
 
\section{ADMR and Quasiparticle residue}
\label{sec:qp_residue}
\subsection{Boltzmann transport approach} 

\begin{figure}
\centering
\includegraphics[width=\columnwidth]{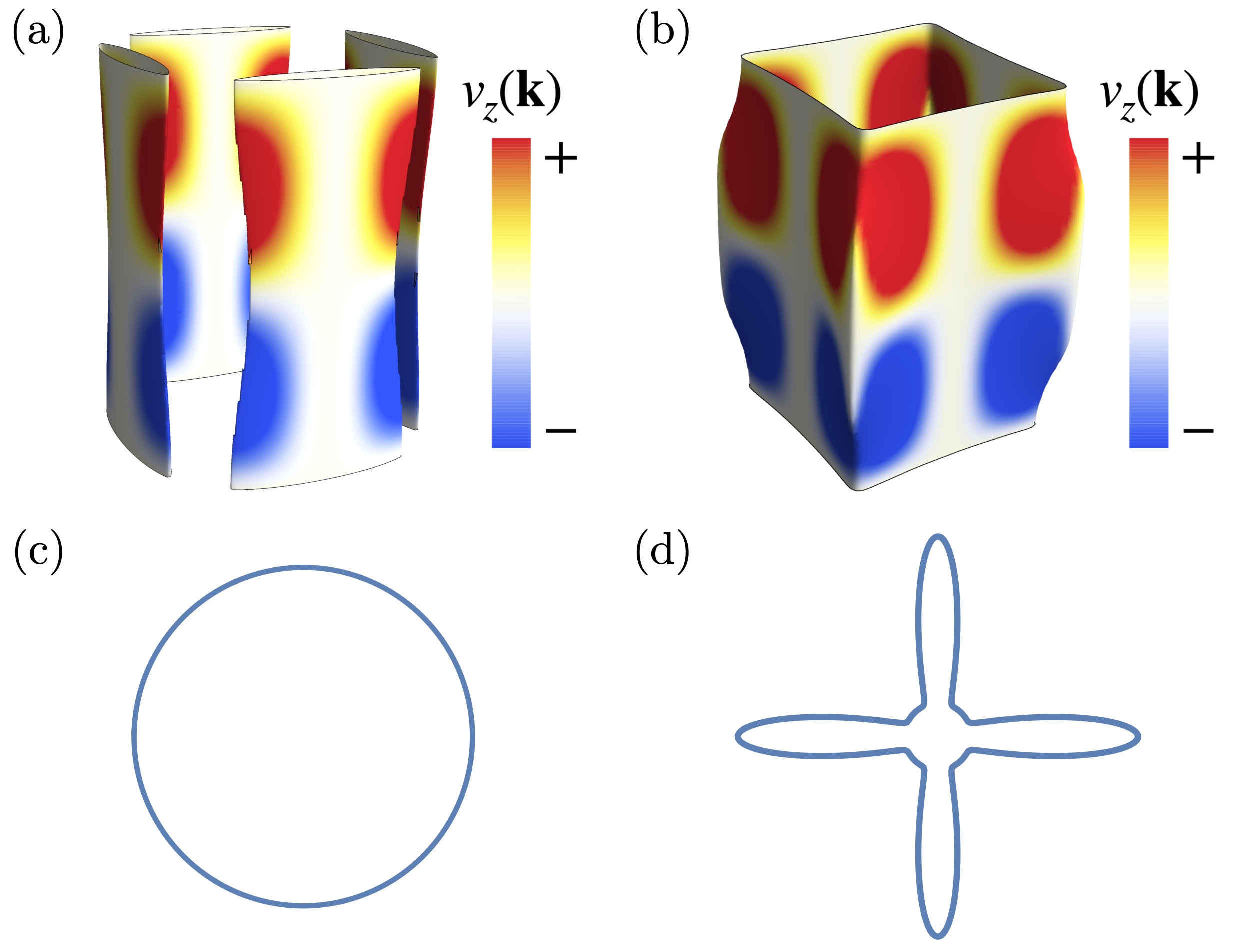}
\caption{A review of the models in \cite{fang_fermi_2020}: (a) their fit to a spin density wave reconstructed Fermi surface for $p=0.21$, where the data and model predictions are shown in \cite{fang_fermi_2020}, (b) their fit to a large Fermi surface at $p=0.24$, (c) their form of $\Gamma = 22.2$THz constant for $p=0.21$, (d) their form of $\Gamma_{\v{k}} = \Gamma_0 + \Gamma_1 \cos^{12}(2\phi)$ with $\Gamma_0 \simeq 15$THz and $\Gamma_1 \simeq 75$THz for $p=0.24$.}
\label{fig:review}
\end{figure}

Consider two dimensional metallic layers oriented in the $xy$-plane and stacked along the $z$-direction. If the in-plane electron dispersion is given by $\epsilon_{\parallel}(\v{k}_\parallel)$, where $\v{k}_\parallel\equiv(k_x,k_y)$, then setting the quasiparticle residue to one, the full dispersion will be given by
\begin{equation}
\epsilon_{\v{k}} = \epsilon_{\parallel}(\v{k}_\parallel) - 2t^\perp_{\v{k}_\parallel} \cos(k_za). \label{eqn:gen_eps}
\end{equation}
Here $t^\perp_{\v{k}_\parallel}$ is the hopping between layers and $a$ is the interlayer separation.\footnote{We will eventually see that if the quasiparticle residue is non-trivial then it will be necessary to renormalize $t^\perp_{\v{k}_\parallel}$ by a multiplicative factor $Z_{\v{k}_\parallel}$.} Note that unlike the toy model proposed in Eq.~\ref{eqn:toy_model_first} the dispersion in-plane is allowed to be arbitrary and the interlayer hopping is allowed to depend on the full in-plane wavevector. This dispersion will lead to a quasi-2d Fermi surface like those pictured in Figure \ref{fig:review}. Within the Boltzmann transport approach if the relaxation-time approximation is employed then the interlayer conductivity $\sigma_{zz}$ is found to be
\begin{equation}
\sigma_{zz} = \frac{e^2}{4\pi^3}\oint \td^2 \v{k} \ \s{D}(\v{k}) v_z(\v{k})\int_{0}^{\infty} \td t \ v_z(\v{k}(t)) e^{-\int_0^t \frac{\td t'}{\tau_{\v{k}(t')}}}\label{eqn:chambers_defn},
\end{equation}
at zero temperature\footnote{This is a good approximation for the experiments done in \cite{fang_fermi_2020}, as $T\lesssim 25K \ll \epsilon_F$ there.}, where the integral is over the Fermi surface, $\s{D}(\v{k})$ is the density of states at $\v{k}$, and the integral in the exponential reflects the probability of scattering after a time $t$. The semi-classical orbits $\v{k}(t)$ are governed by the Lorentz equation $\hbar \dot{\v{k}} = e \v{v}(\v{k})\times \v{B}$ and $\hbar\v{v}(\v{k}) = \boldsymbol{\nabla}_{\v{k}}\epsilon_{\v{k}}$. These orbits are at a constant energy and have a tangent vector perpendicular to $\v{B}$, as seen in Figure \ref{fig:orbits}. 

One can then use Eq.~\ref{eqn:chambers_defn} to go from a model of the Fermi surface and the scattering rate to the angle-dependent magnetoconductance $\sigma_{zz}(\theta',\phi')/\sigma_{zz}(0)$, for a magnetic field of constant magnitude and varying direction: 
\begin{equation}
\v{B} = (B\cos\phi' \sin\theta', B\sin\phi' \sin\theta', B\cos\theta'),
\end{equation}
as displayed in Figure \ref{fig:coord_system}. Since the interlayer conductivity is much smaller than the Hall conductivity in these quasi-2d systems, one can simply invert the angle-dependent magnetoconductance to obtain the angle-dependent magnetoresistance (ADMR) curves $\rho_{zz} \simeq 1/\sigma_{zz}$ \cite{fang_fermi_2020}. Importantly, it is not immediately clear where $Z_{\v{k}_\parallel}$ would enter in the Boltzmann setup, which we address in the next subsection.

\subsection{Decoupled layers approach: addressing $Z_{\v{k}}$}

\begin{figure}
\centering
\includegraphics[width=0.5\textwidth]{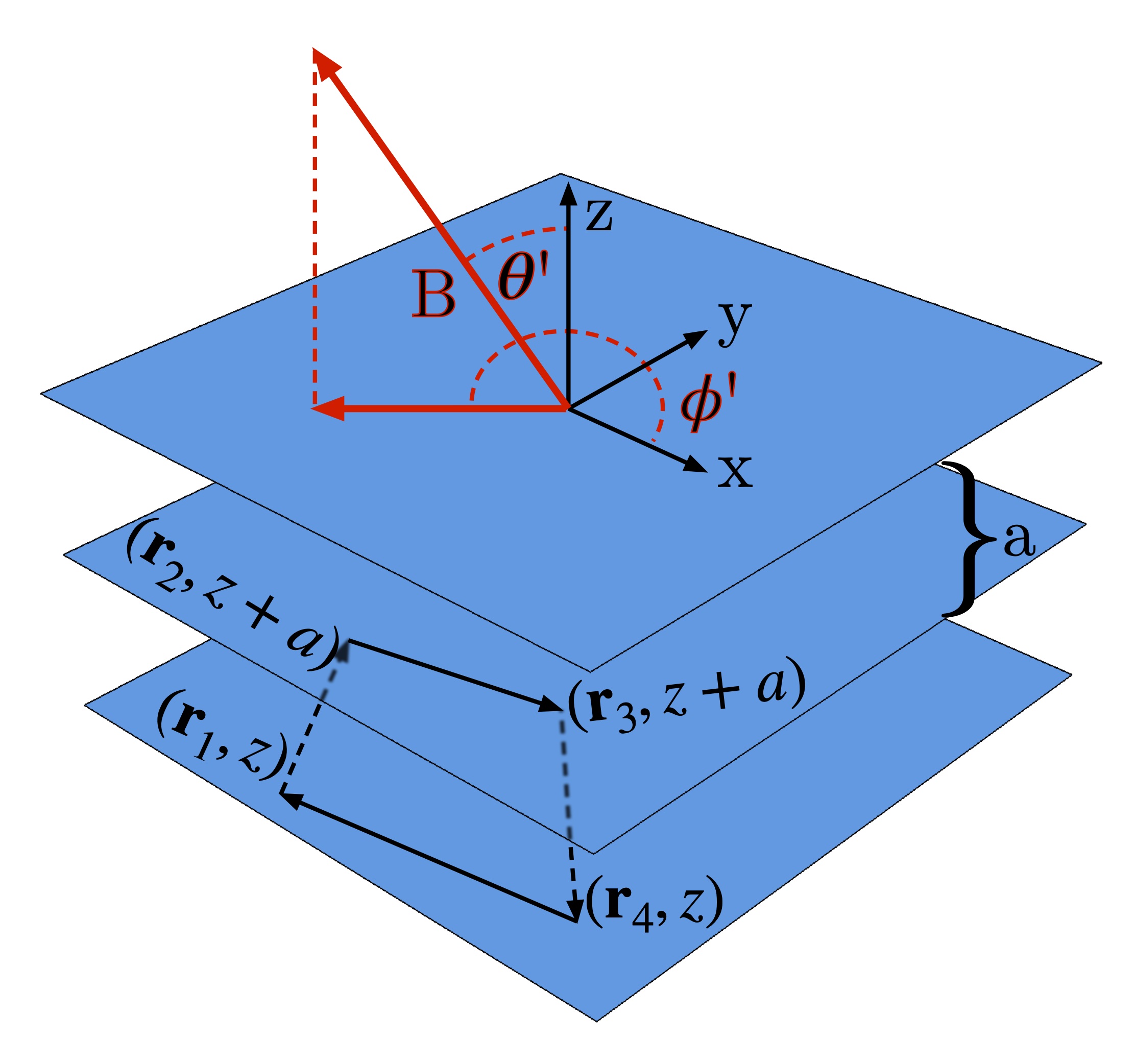}
\caption{Coordinate system used in the Kubo calculation: Fermi liquid layers are shown in blue in real space. Layers are separated by spacing $a$ and coupled to each other by nearest neighbor hopping of magnitude $t_\perp$. The lowest order loop in perturbation theory is shown, which will enclose some flux.}
\label{fig:coord_system}
\end{figure}

Previous work has shown that this Boltzmann transport approach, undertaken in the so-called ``coherent limit" of interlayer transport, is in fact identical to the approach where layers are assumed to be weakly coupled and interlayer transport can be understood as uncorrelated tunneling events, i.e. the ``incoherent limit" \cite{moses_comparison_1999,kennett_sensitivity_2007}. The equivalence of these approaches will hold as long as the $z$-axis conductivity $\sigma_{zz}$ is dominated by the $\s{O}(t^{\perp 2})$ term, i.e. the limit where $t^\perp/\epsilon_F$ is small \cite{kennett_sensitivity_2007}. In the previous work it was suggested that the quasiparticle residue entered by renormalizing the interlayer hopping. Here we address it systematically and verify that $Z_{\v{k}}$ enters via renormalizing the interlayer hopping $t^\perp_{\v{k}}$, $t^\perp_{\v{k}} \rightarrow Z_{\v{k}}t^\perp_{\v{k}}$. As mentioned, we do this via an explicit calculation using the Kubo formula in the small $t^\perp$ limit, shown in Appendix \ref{sect:Kubo_calc}. This calculation begins with uncoupled Fermi liquid layers and couples them perturbatively with a momentum dependent interlayer hopping term. The Kubo formula for interlayer conductivity is then evaluated to one loop order, or order $(t^\perp)^2$, as shown in Figure $\ref{fig:coord_system}$.

We find that the interlayer conductivity is given to this order by:
\begin{align}
\sigma_{zz} \propto& \int \frac{\td^2\v{k}}{(2\pi)^2}\left(t^\perp_{\v{k}}\right)^2 \tilde{A}^B(\v{k}+\v{k}_B/2;\epsilon_F)\tilde{A}^B(\v{k}-\v{k}_B/2;\epsilon_F) \label{eqn:final_Kubo},
\end{align}
where $\v{k}_B = ea (\uv{z}\times \v{B})/\hbar$ and $\tilde{A}^B$ is the gauge invariant piece of the in-plane Fermi liquid's spectral function with a perpendicular magnetic field of $B\cos\theta'$. Physically this can be interpreted as follows. As the particle hops between layers it gets a momentum kick of $\v{k}_B$, arising from the in-plane magnetic field, as is reflected by the momentum difference between the spectral functions. The hopping $t^\perp_{\v{k}}$ can be thought of as occurring between the two layers and thus it acquires only half of the momentum kick. Further the out-of-plane magnetic field will enter via the spectral function $\tilde{A}^B$. In the semi-classical limit this out-of-plane field will induce orbits in-plane that will be reflected as oscillation in $\tilde{A}^B$.

Since $Z_{\v{k}}$ will enter by renormalizing $\tilde{A}^B$ this allows us to see that $t^\perp_{\v{k}} \rightarrow t^\perp_{\v{k}}\sqrt{Z_{\v{k}+\v{k}_B/2}Z_{\v{k}-\v{k}_B/2}}$, where $t^\perp_{\v{k}}$ picks up a factor of $\sqrt{Z}$ from each layer since it is acting between the layers. However, for $Z_\phi=Z_{\v{k}_F(\phi)}$, where $\phi$ is the angle around the Fermi surface, the correction from an added $\v{k}_B$ term will be of order $\v{k}_B\cdot (k_F^{-1}\boldsymbol{\hat{\phi}}\partial_\phi Z_\phi) \sim k_B/k_F$. This quantity will be small compared to the most relevant scale in the problem, the flux through the loop in Figure \ref{fig:coord_system}. For hopping which is local in $\v{r}$, this flux will be on the order of $Bla/l_B^2 \sim k_Bl$ where $l$ is the mean-free path of the particle within a layer. We can thus see that the correction to $Z_{\phi}$ from $\v{k}_B$ will be smaller than this quantity by a factor of $k_Fl$. Since we are assuming the particles are coherent in-plane $k_Fl\gg 1$ and thus we can treat $Z_{\v{k}}$ as simply renormalizing $t^\perp_{\v{k}}$.

As discussed in our summary of results, when we use this result to properly include the quasiparticle residue in the calculation of the ADMR curve with the $(\pi,\pi)$ reconstructed Fermi surface, we find a lack of agreement with the data at $p=0.21$. In light of this it seems as if the striking experimental results from \cite{fang_fermi_2020} may be more interesting, and their interpretation more unresolved, than was initially thought. In order to develop intuition to better understand the experimental data on its own terms, in particular the crossover from upturn on the overdoped side to downturn on the underdoped side, we introduce a simple toy model which we will mine for understanding of these types of behavior. The increased understanding will lead us to propose a simpler explanation for the crossover seen in ADMR data.

\section{Toy model}

We consider a toy model with dispersion given by Eq.~\ref{eqn:toy_model_first}, i.e. a parabolic dispersion in plane with a mass $m^*$ and interlayer separation $a$. For the purpose of discussing the doping of this model we can understand it as the parabolic limit of an in-plane nearest-neighbor hopping model with hopping amplitude $t=\hbar^2/2m^* a_{\parallel}^2$. We will further take the scattering rate to be a constant $\Gamma = 1/\tau$ and assume that $\omega_0\tau = eB\tau/m^*\lesssim 1$. This toy model is chosen to be the simplest possible model which still allows for $\v{k}$-dependent interlayer hopping. Despite its simplicity it will allow us to understand several things about the ADMR curves seen in \cite{fang_fermi_2020}, which we discuss in the summary of our main results.  In this subsection we will not attempt to match the parameters of the model or even the doping level to that directly pertinent to the cuprates. Thus the model in this section should not be viewed as directly relevant to the experiments. Rather our goal will be to explore the physics of what controls whether the ADMR curves show an upturn or downturn within a simple context so as to gain intuition. In the subsequent subsection  we perform calculations with realistic parameters for the cuprates, and the intuition obtained from this subsection will be useful in understanding the results. 

We turn to the specifics of the toy model to elucidate this intuition. First we take $\left(t^\perp_\phi\right)^2$ to have $C_4$ symmetry. As seen in Figure \ref{fig:tperp_van_Hove} the large Fermi surfaces on both sides of the van Hove singularity will possess this symmetry. Then within the semi-classical limit the equations of motion for $\epsilon(k,\phi, k_z)$ are given by $\hbar \dot{\v{k}} = e\v{B}\times \v{v}(\v{k})$, where the full equations of motion can be found as Eq.~\ref{eqn:full_EOM}. To lowest order in $t^\perp$ they are given by $k=\mathrm{const.}, \dot{\phi} = \omega_0\cos\theta'$ and $\dot{k}_z = \omega_0\sin\theta'\sin(\phi-\phi')k$, where $\omega_0 = eB/m^*$.

\begin{figure}
    \centering
    \includegraphics[width=\columnwidth]{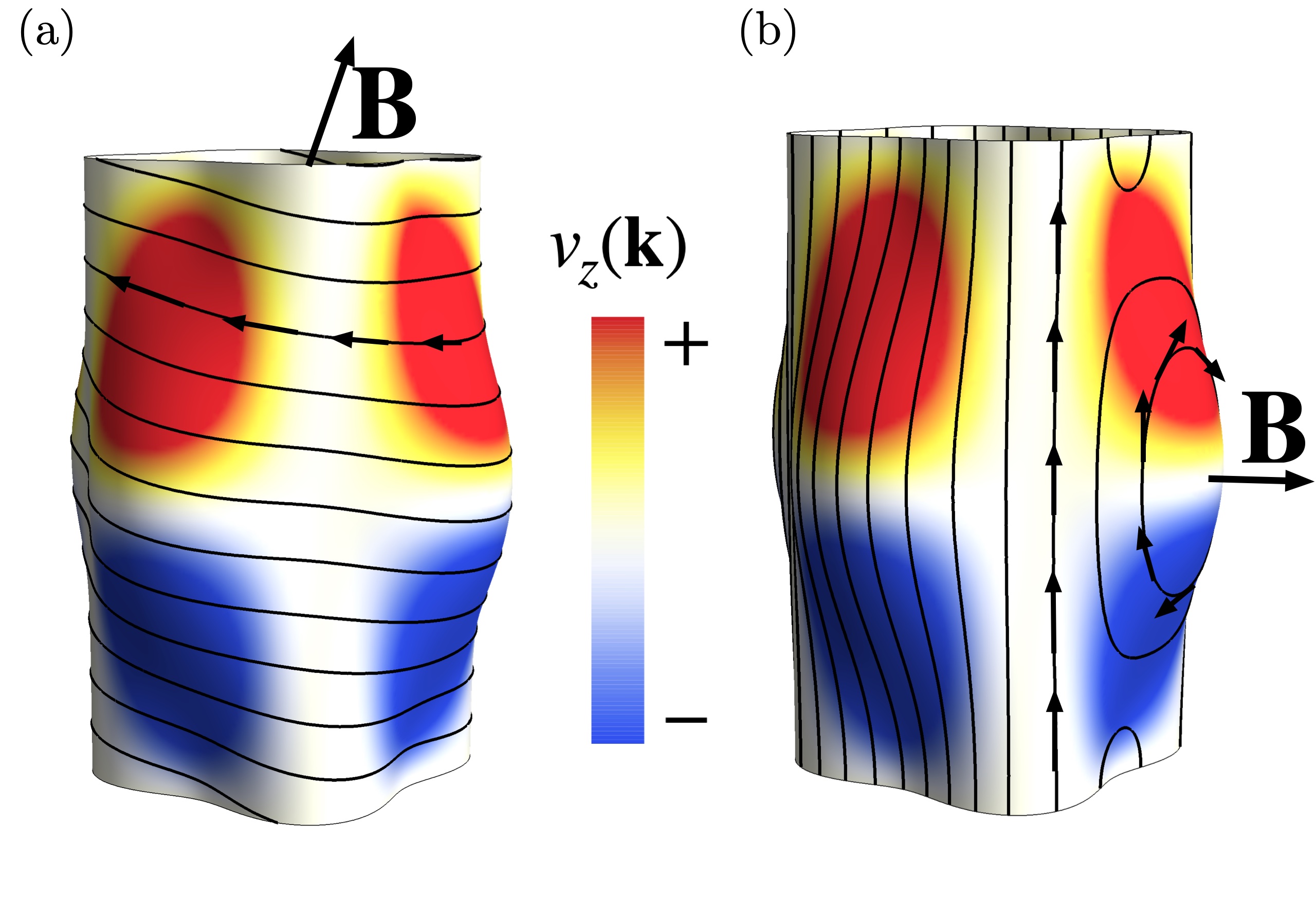}
    \caption{Orbits on a Fermi surface of exaggerated warping. Here $t^\perp_\phi = t_{\parallel}\cos^2(2\phi)/2$, nearly an order of magnitude larger than found in \cite{fang_fermi_2020}, in order to show exaggerated orbits. (a) A magnetic field in a generic direction. Here the orbits all close back on themselves. (b) A magnetic field in-plane. In this case most of the orbits are open. However, some orbits on the ``bulge" of the Fermi surface will be closed. The larger the warping, the larger this bulge is and the more of these closed orbits will be missed by an $\s{O}(t^{\perp 2})$ expansion of Eq.~\ref{eqn:chambers_defn}.}
    \label{fig:orbits}
\end{figure}

In \cite{fang_fermi_2020} the hopping $t^\perp$ is $7\%$ of $\epsilon_F$, so we will consider expanding Chambers' formula Eq.~\ref{eqn:chambers_defn} in powers of $t^\perp/\epsilon_F$. The lowest order term is $\s{O}((t^\perp/\epsilon_F)^2)$, which is all we need. This means dropping all $t^\perp$ terms in $\dot{\v{k}}$ and approximating the Fermi surface as truly quasi-2D, i.e. taking the delta function restriction to the Fermi surface to be $\delta(\hbar^2 k^2 /2m^* - \epsilon_F)$. The trajectories are then
\begin{align}
k =& k_F\\
\phi(t) =& \phi + \omega_0t\cos\theta'\\
k_z(t) =& k_z + k_F\tan\theta'\left[\cos(\phi-\phi') - \cos(\phi(t) -\phi')\right]
\end{align}
The complicated form of $k_z(t)$ reflects the fact that unless $\theta'=90^{\circ}$ the orbits eventually close back on themselves and $k_z$ is periodic as seen in Figure \ref{fig:orbits}(a). Note that if $\theta'=90^{\circ}$ above then $\phi(t) = \phi$ and $k_z(t) = k_z + k_F\omega_0t\sin(\phi-\phi')$; reflecting the open orbits. These are seen in Figure \ref{fig:orbits}(b)

\begin{figure}
\centering
\includegraphics[width=\columnwidth]{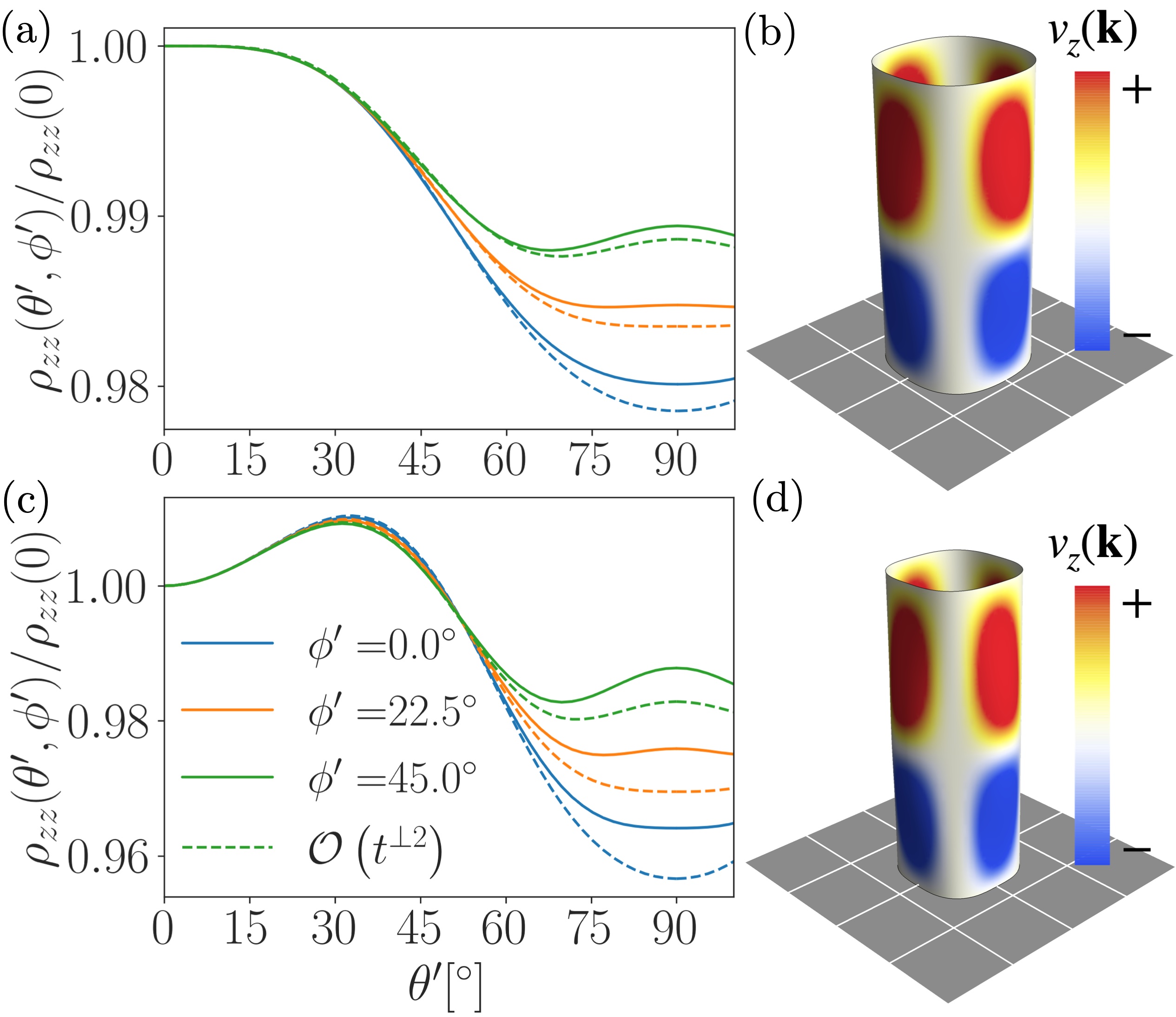}
\caption{Comparison of the $O((t^\perp/\epsilon_F)^2)$ expansion to full Chambers formula. For this model $t^\perp_\phi = t_z\cos^2(2\phi)$ where $t_z = 0.15t_{\parallel}$ and $a_z = 1.76a_{\parallel}$ as in \cite{fang_fermi_2020}. (a) The ADMR curves for $\Gamma = 27.7$THz, $a=1.76a_{\parallel}$, $k_Fa=2.64$, $\omega_0\tau = 0.20$ with the dotted lines tracing out the approximation written as Eq.~\ref{eqn:sig_tsq} and (b) the Fermi surface for this value of $k_F, a$. The colors here display $v_z(\v{k})$. A grid the size of the Brillouin zone was added to make $k_F$ easier to discern. (c) The ADMR curves for $\Gamma = 17.3$THz, $a=1.76a_{\parallel}$, $k_Fa = 2.2$, $\omega_0\tau = 0.32$ with the dotted lines again tracing out the $O(t^{\perp 2})$ expansion and (d) the Fermi surface for this value of $k_F,a$. Note the reduced value of $k_F$ relative to (b).}
\label{fig:tperp_comp}
\end{figure}

Substituting this $O(t^{\perp 2})$ expansion back into Chambers' formula will give us:
\begin{align}
\sigma_{zz}\approx& \frac{e^2 am^*}{\pi^2 \hbar^4}\int \td \phi \ t^\perp_\phi \int_0^\infty \td t \ e^{-t/\tau} t^\perp_{\phi+\omega_0t\cos\theta'}  \label{eqn:sig_tsq}\\ 
\cos&\left\{ak_F\tan\theta'\left[\cos(\phi-\phi') - \cos(\phi-\phi'+\omega_0t\cos\theta')\right]\right\} \nonumber,
\end{align}
where we integrated out the initial $k_z$ dependence.

Figure \ref{fig:tperp_comp} shows a comparison of this approximation with the full Chambers' equation~\ref{eqn:chambers_defn}. The Fermi surfaces in Figure \ref{fig:tperp_comp} were chosen because they generate ADMR curves that resemble the curves seen at both $p=0.24$ and $p=0.21$ in \cite{fang_fermi_2020}, though these Fermi surfaces have unrealistic dopings of $p\approx 0.45$. This is certainly not physically reasonable; our main motivation for persisting with this model for now  is that it will pay dividends by allowing us to understand what exactly drives the upturn vs. downturn behavior at low angles. 

The calculations in Figure \ref{fig:tperp_comp} were computed by using \verb|scikit-image|'s marching cubes algorithm \cite{scikit-image} to discretize the full three-dimensional Fermi suface. Here we have taken $t_z = 0.15t$, a factor of two larger than the value in \cite{fang_fermi_2020}, and nevertheless see that the approximation Eq.~\ref{eqn:sig_tsq} tracks closely with the full Chambers calculation. The greater accuracy of the $\s{O}(t^{\perp2})$ approximation near $\theta'\approx 0$ is because orbits on the warped part of the Fermi surface are heavily dependant on $t^\perp$, but are not excited until $\theta'\approx 90^{\circ}$. This phenomenon can be seen in Figure \ref{fig:orbits}(b) and is discussed by Hanasaki et. al. \cite{hanasaki_contribution_1998} and Schofield and Cooper \cite{schofield_quasilinear_2000} for isotropic interlayer hopping. Schofield and Cooper verified the validity of the $\s{O}(t^{\perp 2})$ expansion for values of $\omega_0\tau$ small enough to not excite these orbits on the warped part of the Fermi surface, provided $t^\perp_\phi =$ const.

In order to understand the behavior in Figure \ref{fig:tperp_comp} reminiscent of the experiments \cite{fang_fermi_2020} we will consider various limits of Eq.~\ref{eqn:sig_tsq}. When the magnetic field is parallel to the layers, i.e. when $\theta'=90^{\circ}$
\begin{align}
\sigma_{zz}\left(90^{\circ},\phi'\right) \propto& \int \td \phi \ \frac{\left(t^\perp_\phi\right)^2}{(\omega_0\tau)^2(k_Fa)^2\sin^2(\phi-\phi') + 1},
\end{align}
where the time integral has been computed exactly. It can already be seen that the dimensionless quantity controlling the peak at $\theta' = 90^{\circ}$ is $\omega_0\tau k_Fa = eaBk_F\tau/m^* = k_Bl$, where $l$ is the mean free path in-plane. As discussed earlier, physically this is the flux, in units of the quantum of flux, through the loop shown in Figure \ref{fig:coord_system} when $\v{r}_1=\v{r}_2, \v{r}_3=\v{r}_4$.

We will spend more time considering low angles, $\theta'\ll 1$. Here we will find that there is no longer a single dimensionless quantity that controls the physics. We first consider $\omega_0\tau$ small enough that all expressions in Eq.~\ref{eqn:sig_tsq} can be expanded. To be more precise, decompose $t^\perp_\phi$ in terms of its harmonics, 
\begin{equation}
t^\perp_\phi = \sum_n c_n e^{in\phi},
\end{equation}
then we will consider the limits $\omega_0\tau k_Fa\ll 1$ and $n_{\mathrm{max}}\omega_0\tau \ll 1$, where $n_{\mathrm{max}}$ is the largest $n$ such that $c_n\neq 0$. We then see that Eq.~\ref{eqn:sig_tsq} at low angles becomes:
\begin{align}
\sigma_{zz} \propto& \int \td \phi \ \left(t^\perp_\phi\right)^2  - (\omega_0\tau)^2 \left[\left(\frac{\partial t^\perp}{\partial \phi}\right)^2 \cos^2 \theta'\right.\\
&\left.+ \left(t^\perp_\phi\right)^2 (k_Fa)^2 \sin^2\theta' \sin^2(\phi-\phi')\right] + \cdots,
\end{align}
where a total derivative of $(t^\perp_\phi)^2$ was discarded and $\partial^2 t^\perp_\phi/\partial \phi^2$ was integrated by parts. The $C_4$ symmetry of $(t^\perp_\phi)^2$ means that when integrated against $\sin^2(\phi-\phi')$ only the constant term will contribute, and we will see that
\begin{align}
\frac{\rho_{zz}(\theta',\phi')}{\rho_{zz}(0)}\approx & 1 + (\omega_0\tau)^2\sin^2\theta' \nonumber \\
&\times \left[\frac{1}{2}(k_Fa)^2 - \frac{\int (\partial t^\perp/\partial \phi)^2 \td \phi}{\int (t^\perp_\phi)^2 \td \phi}\right] \label{eqn:low_ang_low_field}.
\end{align}
This gives us the intuition we discussed in the summary of our main results; for very low values of $\omega_0\tau$ what controls the appearance of an upturn or a downturn at low angles represents a competition between the anisotropy of $t^\perp_\phi$ and the factor $(k_Fa)^2/2$.

However, the situation in Figure \ref{fig:tperp_comp} is not entirely described by this intuition, as there the highest frequency component of $t^\perp_\phi$ is $n_{\mathrm{max}}=4$, so $n_{\mathrm{max}}\omega_0\tau \simeq 1 \simeq \omega_0\tau k_Fa$. We can thus consider the limit of $\omega_0\tau$ small, but not so small that we can expand $t^\perp_\phi$. Now in the case of $\theta'=0$, that is if the magnetic field is perpendicular to the layers, then Eq.~\ref{eqn:sig_tsq} simplifies to 
\begin{align}
\sigma_{zz}(0) \approx& \frac{e^2am^*}{\pi^2 \hbar^4}\int \td \phi \ t^\perp_\phi \int_0^\infty t^\perp_{\phi + \omega_0t}e^{-t/\tau} \ \td t\\
=& \frac{2e^2am^*\tau}{\pi \hbar^4} \sum_n \frac{|c_n|^2}{(n\omega_0 \tau)^2 +1} \label{eqn:sigma_zero},
\end{align}
in terms of the harmonics of $t^\perp_\phi$. We can already see the change that a larger $\omega_0\tau$ will make. The $p$-periodic components of $t^\perp_\phi$ with $p\gg (\omega_0\tau)^{-1}$ will make a vanishingly small contribution to $\sigma_{zz}(0)$, so as $\omega_0\tau$ is increased fewer and fewer components will contribute. This effect arises because $t^\perp_\phi$ is averaged by the in-plane orbits, with higher frequency parts of it being more effectively averaged out. Eventually the only contributing component will be the constant component, if $t^\perp_\phi$ has one, i.e. if it does not average to zero around the Fermi surface. As equation Eq.~\ref{eqn:low_ang_low_field} reveals, a constant $t^\perp_\phi$ will always have an upturn, so this suggests that as $\omega_0\tau$ increases the anisotropy of $t^\perp_\phi$ will get less and less effective at causing a downturn and the system will always achieve an upturn at low angles if $t^\perp_\phi$ does not average to zero. 

We can check this intuition by going to the next order in $\omega_0\tau k_Fa\sin\theta'$ where we will have
\begin{align}
&\sigma_{zz}(\theta',\phi') \propto \left[\sum_{n}\frac{|c_{n}|^2}{(n\omega_0\tau \cos\theta')^2 +1}- \frac{1}{2}(k_Fa)^2(\omega_0\tau)^2\right. \nn
&\left.\times \sin^2\theta'\left(\sum_{n}|c_{n}|^2 \frac{1-3(n\omega_0\tau \cos\theta')^2}{[(n\omega_0\tau \cos\theta')^2 + 1]^3}\right)\right] \label{eqn:sigma_low_ang},
\end{align}
and where we have assumed that $\omega_0\tau\lesssim 1$. This expression is somewhat opaque but can be simplified by considering two cases. In the first, as in Figure \ref{fig:tperp_comp}, $t^\perp_\phi$ has a constant component and does not average to zero across the Fermi surface. In the second case it will. In the first case as $\omega_0\tau$ increases $\sigma_{zz} \sim |c_0|^2 - |c_0|^2(\omega_0\tau k_Fa)^2\sin^2\theta'/2$ and there will always be an upturn in resistivity at low angles as our intuition predicted. 

The intuition behind the case when $\langle t^\perp_\phi\rangle \neq 0$ allows us to describe Figure \ref{fig:tperp_comp} more completely. In both Figures \ref{fig:tperp_comp}(a) and (c), $k_Fa$ was chosen so that $(k_Fa)^2/2 < \int (\partial t^\perp/\partial \phi)^2 \td \phi/ \int (t^\perp_\phi)^2 \td \phi$ so that there was a downturn for very small $\omega_0\tau$. The value of $\omega_0\tau$ was then increased, and since both systems' $t^\perp_\phi$ did not average to zero this was guaranteed to lead to an upturn. In Figure \ref{fig:tperp_comp}(c) $\omega_0\tau = 0.32$ which is big enough that the system thus has an upturn at low angles, while in Figure \ref{fig:tperp_comp}(a) $\omega_0\tau$ is small enough that this does not occur. In order to nevertheless get an upturn at $\theta'=90^{\circ}$ of the same magnitude, which is controlled by $\omega_0\tau k_Fa$, Figure \ref{fig:tperp_comp}(a) was chosen to have a larger value of $k_Fa$. Thus as long as $\langle t^\perp_\phi \rangle \neq 0$ one can always tune $\omega_0\tau$ to show an upturn at low angles, while also tuning $k_Fa$ so $\omega_0\tau k_Fa$ is large enough to show an upturn when $\theta'=90^{\circ}$.

\begin{figure}
\centering
\includegraphics[width=\columnwidth]{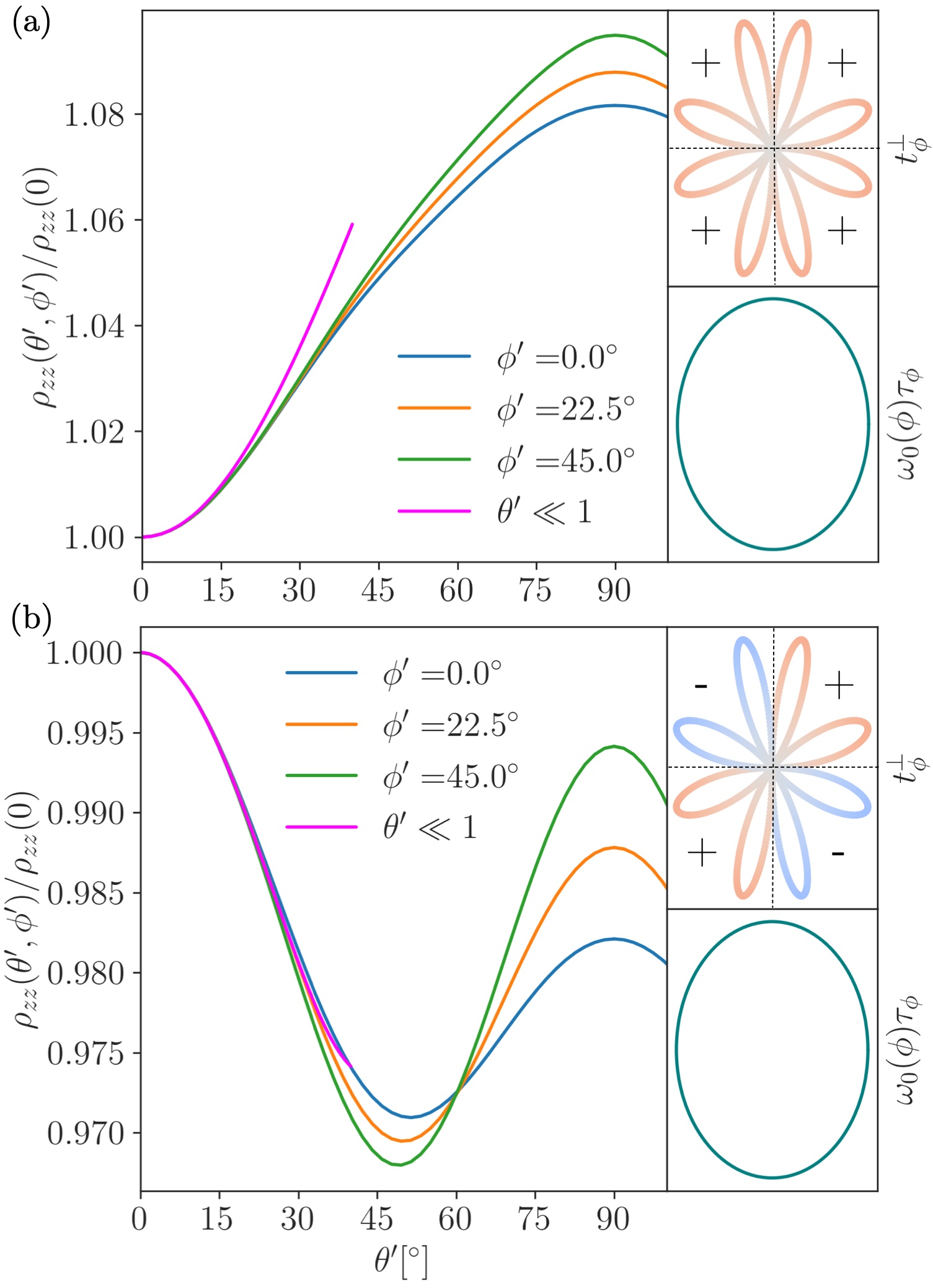}
\caption{Change in behavior at low angles when $\langle t^\perp_\phi \rangle = 0$. Both plots consider the toy model and have constant $\omega_0(\phi)\tau_\phi = 0.125$, shown in the inset, and $k_Fa = 6$. (a) Here $t^\perp_\phi = |\sin(2\phi)|\cos^2(2\phi)$ and is everywhere positive, as is displayed in the inset. The system thus has an upturn for $\omega_0\tau$ large enough, despite having a downturn at very low $\omega_0\tau$. Note also that Eq.~\ref{eqn:sigma_low_ang}, plotted in magenta, does agree well at low angles. (b) The only change here is $t^\perp_\phi = \sin(2\phi)\cos^2(2\phi)$, which therefore changes sign in each quadrant, as displayed in the inset. Since $t^\perp_\phi$ averages to zero we expect a downturn, which is indeed obtained.}
\label{fig:tperp_sign}
\end{figure}

The situation is dramatically different if $t^\perp_\phi$ does indeed average to zero. In this case the lowest order $|c_0|^2$ terms no longer contribute. Then the lowest order nonzero term for small $\theta'$ is the term $|c_{n_{\mathrm{min}}}|^2/((n_{\mathrm{min}}\omega_0\tau\cos\theta')^2+1)$ where $n_{\mathrm{min}}$ is the lowest nonzero frequency component of $t^\perp_\phi$. This will lead to an ADMR curve of the form
\begin{equation}
\frac{\rho_{zz}(\theta',\phi')}{\rho_{zz}(0)}\sim \frac{(n_{\mathrm{min}}\omega_0\tau\cos\theta')^2+1}{(n_{\mathrm{min}}\omega_0\tau)^2+1},
\end{equation}
where $n_{\mathrm{min}}\omega_0\tau = 0.25$ for Figure \ref{fig:tperp_sign}(b). This form of the ADMR curve will always display a downturn in resistance as the magnetic field is swept parallel to the layers, provided $\omega_0\tau$ is large enough. Figure \ref{fig:tperp_sign} illustrates the difference between the cases when $\langle t^\perp_\phi\rangle \neq 0$ and when $\langle t^\perp_\phi \rangle =0$.

The downturn when $\langle t^\perp_\phi \rangle = 0$ can be intuitively understood as arising from the
\begin{equation}
\int \td \phi \ t^\perp_\phi \int_0^\infty t^\perp_{\phi+\omega_0t\cos\theta'}e^{-t/\tau}
\end{equation}
term in the conductivity. For $\omega_0\tau$ large enough the time integral will begin to average $t^\perp$ to zero and the conductivity will thus be made very small. Tilting the magnetic field decreases the distance that the electrons travel in-plane before scattering by a factor of $\cos\theta'$ and thus makes the averaging less effective. This will increase the conductivity and lead to a downturn in the ADMR curve.

\section{Results on models for the cuprates} 

\begin{figure}
\centering
\includegraphics[width=\columnwidth]{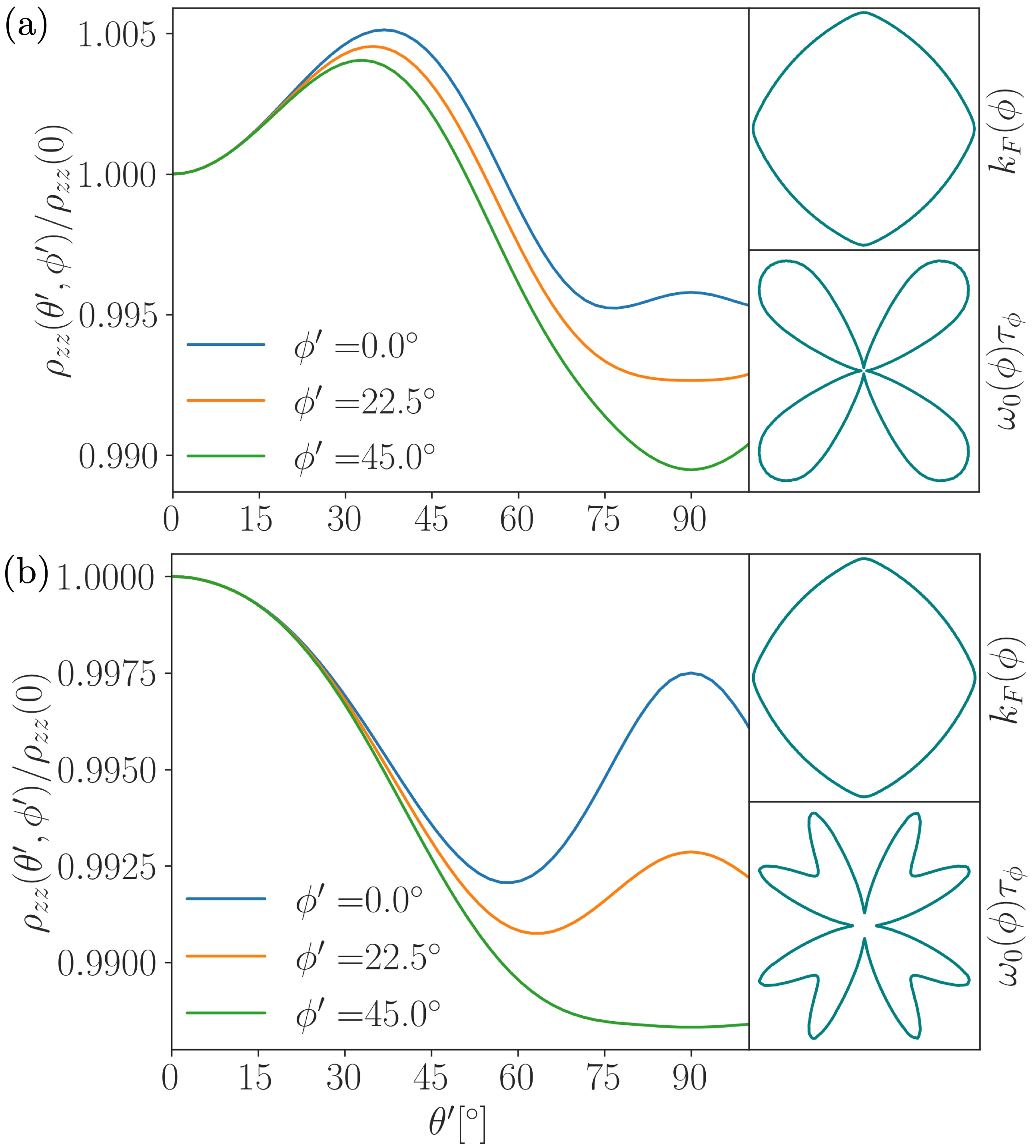}
\caption{Behavior of ADMR curves from large Fermi surface upon renormalizing $\v{v}_F(\phi)$. (a) Here the same parameters are used as in \cite{fang_fermi_2020}, with the exception of the chemical potential which is tuned  to give a doping of $p=0.210 \pm 0.003$. This results in a Fermi surface centered at $(\pi,\pi)$ with a $k_F(\phi)$ shown in the inset. The curve displays an upturn at low angles, but this can be attributed to the smallness of $\omega_0(\phi)\tau_\phi$ at the antinode, shown in the inset. The average of $\omega_0(\phi)\tau_\phi$ over the full Fermi surface is $\langle \omega_0(\phi)\tau_\phi \rangle = 0.093$ in this case. (b) This ADMR curve employs the same parameters, but with the Fermi velocity renormalized by a factor of $(1+4\cos^2(2\phi))$ and the scattering rate now proportional to $1+5\cos^{12}(2\phi)/2$ rather than $1+5\cos^{12}(2\phi)$. This results in a $\omega_0(\phi)\tau_\phi$ which is substantially increased near the antinodes, shown in the inset. Again $\langle \omega_0(\phi)\tau_\phi \rangle = 0.093$ across the Fermi surface, where the overall scattering rate was increased to compensate for the increased $\omega_0(\phi)$ near the antinode. For this ADMR curve the predicted downturn at low angles is captured.}
\label{fig:v_renorm}
\end{figure}

As we suggested when introducing our toy model, the fact that a change from $\langle t^\perp_\phi \rangle \neq 0$ to $\langle t^\perp_\phi \rangle =0$ can change an upturn at low angles to a downturn suggests that the transition seen in \cite{fang_fermi_2020} is actually driven by this type of physics. As we mentioned in our summary of results, such a transition certainly does occur between $p=0.24$ and $p=0.21$ at temperatures above the pseudogap temperature, driven by the change from a large Fermi surface centered at the $\Gamma$-point on the overdoped side to one centered at $(\pi,\pi)$ on the underdoped side and the form of the hopping $t^\perp_{\v{k}}$ Eq.~\ref{eqn:tperp_k_form}. A test of this theory is shown in Figure \ref{fig:v_renorm}(a), which shows that the curves at doping $p=0.21$ still have an upturn at low angles despite $t^\perp_\phi$ now averaging to zero.

This continued upturn at low angles is belied by the very anisotropic $\omega_0(\phi)\tau_\phi$ around the Fermi surface, displayed in an inset, where 
\begin{equation}
\omega_0(\phi) = eB\frac{\v{v}_F(\phi)\cdot \v{k}_F(\phi)}{\hbar k_F(\phi)^2}
\end{equation}
for a two-dimensional Fermi surface with angle dependent $k_F(\phi)$ and $v_F(\phi)$ \cite{kennett_sensitivity_2007}. Here $\v{k}_F(\phi) = (k_F(\phi)\cos\phi, k_F(\phi)\sin\phi)$ by definition. As we saw from the toy model the crucial reason for the sharp downturn appearing as $\omega_0\tau$ is increased is that $t^\perp_\phi$ is more effectively averaged by the in-plane orbits. If $\omega_0(\phi)\tau_\phi$ is very small at the antinodes, where the change in sign of $t^\perp_\phi$ occurs, this will mean the quasiparticles do not visit areas of the Fermi surface with a different sign of $t^\perp_\phi$ in their orbits, hindering the ability of the system to average $t^\perp_\phi$. 

The velocity $\v{v}_F(\phi)$ can nonetheless be renormalized across $p^*$, increasing $\omega_0(\phi)\tau_\phi$ at the antinodes. This is shown in Figure \ref{fig:v_renorm}(b) for the renormalization of the Fermi velocity $\omega_0(\phi)\rightarrow (1+4\cos^2(2\phi))\omega_0(\phi)$ and a scattering rate now proportional to $1+5\cos^{12}(2\phi)/2$ rather than $1+5\cos^{12}(2\phi)$ \ \footnote{The angle $\phi$ for the scattering rate is measured with respect to $(0,0)$, with all coordinates being translated to the first Brillouin zone in order to compute it. This is done in order to be consistent with the approach taken in \cite{fang_fermi_2020}. For all other quantities the angle $\phi$ references the angle around the large Fermi surface and is thus measured with respect to $(\pi/a_{\parallel},\pi/a_{\parallel})$}. With this renormalization $t^\perp_\phi$ can be averaged to zero and the initial upturn at low $\theta'$ disappears. Our choices for this particular renormalization and scattering rate illustrate that a $\phi'=0$ curve can be obtained which is similar in amplitude and form to the $p=0.21$ data displayed in \cite{fang_fermi_2020}. Importantly the appearance of this downturn persisted for a number of scenarios that renormalized the Fermi velocity to have a larger value at the antinodes, including $\omega_0(\phi)\sim \mathrm{const.}$ or $\omega_0(\phi)\sim k_F^{-1}(\phi)$ with and without a change to the scattering rate. In all of these scenarios the value of $\omega_0(\phi)\tau_\phi$ was still larger in the nodal region than the antinodal region, the anisotropy was merely lessened. We thus conclude that as long as the Fermi velocity is renormalized to have a reasonably larger value at the antinodes than that suggested by the band theory the downturn at low angles for $p=0.21$ can be explained by averaging over a $t^\perp_\phi$ that now averages to zero around the large Fermi surface.

There is still some discrepancy with the $\phi'$ dependence at $\theta'=90^{\circ}$, namely there is less broadening at $\theta'=90^{\circ}$ with respect to $\phi'$ in the $p=0.21$ data than in Figure \ref{fig:v_renorm}(b). We speculate that this is due to the relative lack of warping in the $k_z$ direction of the large Fermi surface. We have found numerically that the peak at $\theta'=90^{\circ}$ is strongly controlled by this warping in the case of the $(\pi,\pi)$ reconstructed hole pockets. The $k_z$ warping is an effect beyond $\s{O}(t^{\perp 2})$ in the Chambers' integral, and we can verify that the $\s{O}(t^{\perp 2})$ approximation leads to significant broadening at $\theta'=90^{\circ}$. Indeed, an $\s{O}(t^{\perp 2})$ treatment of $\sigma_{zz}$ for a Fermi surface with a varying $k_F(\phi), \omega_0(\phi), \tau_\phi$ within linear response (see Appendix \ref{sec:lin_resp_parallel}) is found to give
\begin{equation}
\sigma_{zz}(90^{\circ},\phi') \sim \int \frac{\td \phi}{|\v{v}_F(\phi)|} \frac{\left(t^\perp_\phi\right)^2 \tau_\phi}{1+(\tau_\phi \v{v}_F(\phi)\cdot \v{k}_B)^2},
\end{equation}
where $\v{k}_B = \frac{ea}{\hbar}\v{B}\times \uv{z}$ and can be thought of as the momentum kick the particle receives from the magnetic field when it hops from layer to layer. The term $\tau_\phi \v{v}_F(\phi)\cdot \v{k}_B$ can be reduced to
\begin{equation}
\omega_0(\phi)\tau_\phi k_F(\phi)a\left(\sin(\phi-\phi')+\frac{k_F'(\phi)}{k_F(\phi)}\cos(\phi-\phi')\right),
\end{equation}
which means that if $k_F(\phi)=k_F$ and the other quantities are tuned so that 
\begin{equation}
\frac{\left(t^\perp_\phi\right)^2 \tau_\phi (\omega_0(\phi)\tau_\phi k_Fa)^4}{\omega_0(\phi)k_F}
\end{equation}
has no period-$4$ component, then broadening of the ADMR curves with respect to $\phi'$ will not take place until $\s{O}((\omega_0\tau k_Fa)^8)$ significantly lessening the broadening. However, if $k_F$ varies with $\phi$, as in the large Fermi surface, the additional $\cos(\phi'-\phi)$ term in the integrand will make it very difficult to push the broadening out to this order.

Thus we suspect that this discrepancy of broadening at $\theta'=90^{\circ}$ might be resolved by  scanning the parameter space of warping in the $k_z$ direction. We will however not pursue this here. Nevertheless, we have demonstrated that by renormalizing the Fermi velocity the downturn that is seen in the $p=0.21$ data can be obtained simply by averaging over $t^\perp_\phi$, which averages to zero on the underdoped side. This represents a simpler explanation for the crossover seen in the ADMR data in \cite{fang_fermi_2020} than a $(\pi,\pi)$ reconstruction in-plane.

\section{Discussion}
In this paper, inspired by the striking experimental results of \cite{fang_fermi_2020} we developed a framework for interpreting ADMR measurements in layered correlated metals. Specifically, we argue based on a careful analysis that the effect of the quasiparticle residue $Z_\v{k}$ and its momentum dependence can impact the calculation of the $c$-axis transport in significant ways. By studying a toy model, we developed some intuition on what controls whether at low tilt angles the ADMR curves show an upturn or downturn, as seen in \cite{fang_fermi_2020} when going from doping $p=0.24$ to doping $p=0.21$. We proposed that this change is caused by the change of the large Fermi surface as it crosses a van-Hove point. 

A well-informed reader might object that our proposal ignores the body of experimental literature showing that the carrier density in Nd-LSCO drops precipitously when $p$ decreases to $0.21$ from $0.24$. This is demonstrated for instance by Hall transport experiments. However we remark that the pseudogap scale in Nd-LSCO in this doping range is actually quite low (of order $50-70 K$) while the ADMR experiments only go down to $25 K$. The drastic change of Hall number happens at much lower temperature. Thus thinking in terms of the large Fermi surface may not be unreasonable for the purposes of understanding the ADMR results, as they currently stand. Though the large Fermi surface below the critical doping cannot explain every detail of the data, such as features occurring when the magnetic field is rotated in-plane, it provides a simple explanation for this important change. Further, our detailed calculations (and importantly the inclusion of the angle-dependent $Z$ caused by the folding of the Brillouin zone) do not support the possibility that a $(\pi,\pi)$ modulated reconstruction reproduces the measured ADMR curve at $p = 0.21$. This is consistent with the lack of any independent evidence for such order at that doping and at the temperature scale of the ADMR measurement. We cannot, however, rule out more exotic scenarios with more free paramaters such as FL* or its particular case, the YRZ ansatz \cite{yang_phenomenological_2006,moon_underdoped_2011}.

Another potential objection to our analysis (as well as all prior discussions) of ADMR is that we have used a quasiparticle/Boltzmann equation description while much of the data (for $p = 0.24$) is in a regime where the $ab$-plane resistivity is linear in $T$, and hence in the `strange metal' regime. Indeed analysis of the same ADMR data at $p = 0.24$ in \cite{grissonnanche_linear-temperature_2021} within the Boltzmann transport framework lead to a `near-Planckian' transport scattering scattering rate $\Gamma_{ab} = \alpha \frac{k_BT}{\hbar} $ with $\alpha = 1/2 \pm 0.4$. There are two reasons why our analysis may be legitimate in this regime. As noted in a number of prior papers, the Boltzmann approach can be justified \cite{prange_transport_1964} so long as the electron self-energy depends only on frequency and not on momentum - in the cuprates it seems to be the case that the self-energy is mostly independent of the momentum normal to the Fermi surface at low energies \cite{sobota_angle-resolved_2021}. A second observation is that the in-plane scattering rate $\Gamma_{ab}$ is comparable (or smaller) than the inter-plane hopping rate $t^\perp_{\mathbf k_\parallel}$.  Thus the electron may retain its integrity  on the time scale required to hop between two adjacent layers, and this may be enough to contribute to the ADMR signal. However we leave a full understanding of the effect of loss of quasiparticle coherence on ADMR to the future.

To conclude, let us return to the question of how the electronic Fermi surface evolves from overdoped to underdoped at $T = 0$ (accessed by suppressing superconductivity in a magnetic field) across a putative quantum critical point. Directly probing this through ADMR will require experiments at even lower temperatures. This is complicated by the need to tilt the field which reduces the out-of-plane component and hence the efficiency with which the superconductivity may be suppressed. Other indirect experiments such as studies of quasiparticle interference using scanning tunneling microscopy may perhaps also provide useful information. 

\acknowledgements
We thank Brad Ramshaw, Ga{\"e}l Grissonnanche and Yawen Fang for innumerable conversations that played a crucial role in shaping our thinking. We also thank Alexander Nikolaenko for his critical reading of our manuscript. SM was supported by the National Science Foundation Graduate Research Fellowship under Grant No. 1745302. Any opinions, findings, and conclusions or recommendations expressed in this material are those of the author(s) and do not necessarily reflect the views of the National Science Foundation.  DC was supported by a faculty startup grant from Cornell University. PL acknowledges support by US Department of Energy office of Basic Sciences grant number DE-FG02-03ER46076. TS was supported by US Department of Energy grant DE- SC0008739, and partially through a Simons Investigator Award from the Simons Foundation.  This work was also supported by the Simons Collaboration on Ultra-Quantum Matter, which is a grant from the Simons Foundation (651446, TS).

\appendix

\section{ADMR in linear response} 
\label{sect:Kubo_calc}

Consider a stack of two-dimensional Fermi liquids with the $z$-direction along the normal and $\v{r} = (x,y)$ labeling the position of points in a plane. Let $a$ be the lattice spacing, and $z_n = na$ be the position of the $n-$th plane. We denote $c_{\v{r},z_n}$ as the annihilation operator for an electron at $\v{r}$ in the $n-$th plane. Further, let $\v{A} = (A_x, A_y)$ and $A_z$ together make up the vector potential, in order to emphasize the separation of the $z$ direction from the others by translation-symmetry breaking. This situation is shown in Figure \ref{fig:coord_system}. If we define the vector potential as
\begin{equation}
\v{A} = B\left(x\cos\theta' \uv{y} - x\sin\theta'\sin\phi'\uv{z} + y\sin\theta'\cos\phi'\uv{z}\right)
\end{equation}
then if the action in the $n$th Fermi liquid layer is written as $S_n[\v{A}]$ the full action can be written as 
\begin{align}
S[\v{A},A_z] =& \sum_n S_n[\v{A}] + S_{\mathrm{hop}}[\v{A},A_z], \text{ where} \label{eqn:action}\\
S_{\mathrm{hop}}[\v{A},A_z] =& -\int \td \tau \sum_{\v{r}_1,\v{r}_2,z} e^{i\theta_B(\v{r}_2, z+a,\v{r}_1,z;\tau)}\nn
&\times t^\perp(\v{r}_2-\v{r}_1)c^\dagger_{\v{r}_2,z+a}c_{\v{r}_1,z} + \mathrm{h.c.}\\
\theta_B(\v{r}_2,z+a,\v{r}_1,z) =& \int_{(\v{r}_1,z)}^{(\v{r}_2,z+a)} \v{A}\cdot \td \v{l}\nn
=& \frac{B\cos\theta'}{2}(x_1+x_2)(y_2-y_1)+ Ba\sin\theta' \nn
&\times \left[\frac{y_1+y_2}{2}\cos\phi' - \frac{x_2+x_2}{2}\sin\phi'\right],
\end{align}
where this latter term is equivalent to $aA_z\left(\frac{\v{r}_1+\v{r}_2}{2};\tau\right)$ and where $t^\perp(\v{r})$ is the Fourier transform of $t^\perp_{\v{k}}$.

Within linear response the conductivity is given by 
\begin{equation}
\sigma_{ij}(\v{k}, \omega) =  \frac{\beta}{V}\frac{1}{i\omega}\frac{\delta^2\ln\left(\s{Z}[\v{A},A_z]\right)}{\delta A_j(\v{k},\omega)\delta A_i(-\v{k},-\omega)}\label{eqn:sigmazz_linresp},
\end{equation}
where $V$ is the three-dimensional volume of the sample and where $\s{Z}$ is the partition function of Eq.~\ref{eqn:action}. The expression $\ln(\s{Z})$ can be expanded perturbatively in powers of $t^\perp/\epsilon_F$. To the lowest order this will be given by
\begin{align}
\ln(\s{Z}) =& \int \td\tau\td \tau' \sum_{\v{r}_1,\ldots,\v{r}_4,z} t^\perp(\v{r}_2-\v{r}_1)t^\perp(\v{r}_4-\v{r}_3)\nn
&\times e^{ie[\theta_B(\v{r}_2,z+a,\v{r}_1,z;\tau)-\theta_B(\v{r}_3,z+a,\v{r}_4,z;\tau')]}\nn
&\times G^B(\v{r}_2,\v{r}_3;\tau,\tau')G^B(\v{r}_4,\v{r}_1;\tau'\tau),
\end{align}
where $G^B(\v{r},\v{r}';\tau,\tau')$ is the Green's function for propagation from $\v{r},\tau$ to $\v{r}',\tau'$ within a single uncoupled layer with perpendicular magnetic field of $B\cos\theta'$.

We now define $K(i\omega_n) = [i\omega\sigma(\v{k}=0,\omega)]_{\omega\rightarrow i\omega_n}$. We can then take the functional derivatives of Eq.~\ref{eqn:sigmazz_linresp} and transform the Green's functions to Matsubara frequency to arrive at 
\begin{align}
K(i\omega_n) =& \frac{N(ea)^2}{\beta V}\sum_{\omega_k}\sum_{\v{r}_1,\ldots,\v{r}_4}  t^\perp(\v{r}_2-\v{r}_1)t^\perp(\v{r}_4-\v{r}_3)\nn
&\times e^{ie[\theta_B(\v{r}_2,a,\v{r}_1,0;\tau)-\theta_B(\v{r}_3,a,\v{r}_4,0;\tau')]}\s{G}^B(\v{r}_2,\v{r}_3;i\omega_k)\nn
&\times \left[\s{G}^B(\v{r}_4,\v{r}_1; i\omega_k + i\omega_n)+\s{G}^B(\v{r}_4,\v{r}_1; i\omega_k- i\omega_n)\right.\nn
&\left.-2\s{G}^B(\v{r}_4,\v{r}_1; i\omega_k)\right] \label{eqn:matsub_sum},
\end{align}
where $N$ is the number of layers and we used the fact that $\theta_B$ has no $z$ dependence to sum over $z$. These Matsubara Green's functions will have the form 
\begin{equation}
\s{G}^B(\v{r},\v{r}';i\omega_n) = g^B(\v{r}'-\v{r};i\omega_n)e^{i\theta_B(\v{r},\v{r}')},
\end{equation}
where we write $\theta_B(\v{r},\v{r}')$ for the phase picked up traveling from $\v{r}$ to $\v{r}'$ in-plane and where $g^B$ will be a gauge invariant function. The spectral function $A^B$ defined by the relation
\begin{equation}
\s{G}^B(\v{r},\v{r}';i\omega_n) = \int \td \Omega \frac{A^B(\v{r},\v{r}';\Omega)}{i\omega_n - \Omega},
\end{equation}
can then be shown to have the form $A^B(\v{r}, \v{r}';\omega) = e^{i\theta_B(\v{r},\v{r}')}\mathrm{Im}(g^B(\v{r}'-\v{r};\omega))/\pi$. If $\tilde{A}^B$ is taken to be the gauge invariant piece of the spectral function, then it can be defined as
\begin{equation}
g^B(\v{r}'-\v{r};i\omega_n) = \int \td \Omega \frac{\tilde{A}^B(\v{r}'-\v{r};\Omega)}{i\omega_n - \Omega}.
\end{equation}

With these definitions $g^B$ can be substituted in place of $\s{G}$ in Eq.~\ref{eqn:matsub_sum}, these can be transformed to $\tilde{A}^B$, and the resulting Matsubara sum can be completed. The resulting DC conductivity will be given by
\begin{align}
\sigma_{zz} =& \frac{2\pi N(ea)^2}{V}\int \td \Omega \sum_{\v{r}_1,\ldots,\v{r}_4} e^{i\Phi}t^\perp(\v{r}_2-\v{r}_1)t^\perp(\v{r}_4-\v{r}_3)\nn
&\left(-\frac{\partial f}{\partial \Omega}\right)\tilde{A}^B(\v{r}_3-\v{r}_2;\Omega)\tilde{A}^B(\v{r}_1-\v{r}_4;\Omega) \label{eqn:Kubo_real},
\end{align}
where $\Phi$ is the flux through the loop shown in Figure \ref{fig:coord_system} and we can approximate $-\partial f/\partial \Omega = \delta(\Omega-\epsilon_F)$. This formula reproduces the familiar Kubo formula for conductivity if $t^\perp_{\v{k}} = t^\perp$ and is manifestly gauge invariant. As noted earlier, this equation is equivalent to the Chambers' formula to $\s{O}(t^{\perp 2})$ in the semi-classical limit \cite{kennett_sensitivity_2007}. 

The equation Eq.~\ref{eqn:Kubo_real} is invariant under translation of the loop shown in Figure \ref{fig:coord_system} by any constant vector and thus we freely set $\v{r}_1$ to occur at the origin and complete one sum over the coordinates. This will then reduce to 
\begin{align}
\sigma_{zz} =& 2\pi ae^2 \sum_{\v{d}_1,\v{d}_2,\v{d}_3} e^{i\Phi(\v{d}_1,\v{d}_2,\v{d}_3)}\tilde{A}^B(\v{d}_3;\epsilon_F)\tilde{A}^B(\v{d}_1;\epsilon_F)\nn
&\times t^\perp(\v{d}_2)t^\perp(-\v{d}_1-\v{d}_2-\v{d}_3)\label{eqn:Kubo_change},
\end{align}
under a coordinate change $\v{d}_i = \v{r}_i - \v{r}_{i-1}$ for $i=1,2,3,4$ and where $\v{r}_0=\v{r}_4$ by definition. In these coordinates the flux through the loop in Figure \ref{fig:coord_system} will be given by
\begin{align}
\Phi(\v{d}_1,\v{d}_2,\v{d}_3) =& e\v{B} \cdot \frac{1}{2}\left[(\v{d}_1+\v{d}_2,a)\times (\v{d}_2+\v{d}_3,a)\right]\nn
=& e\v{B}\cdot \frac{1}{2}\left[(\v{d}_1+\v{d}_2)\times (\v{d}_2+\v{d}_3)\right]\nn
&+ \frac{1}{2}\v{k}_B \cdot (\v{d}_1-\v{d}_3),
\end{align}
where $\v{k}_B = ea(\uv{z}\times \v{B})$ can be thought of as a momentum kick the particle receives between layers.

If Eq.~\ref{eqn:Kubo_change} is then partially Fourier transformed it will be given by
\begin{align}
\sigma_{zz} =& 2\pi ae^2\sum_{\v{d}_1,\v{d}_3}\int \frac{\td^2 \v{k}}{(2\pi)^2} e^{i\v{k}_B \cdot (\v{d}_1-\v{d}_3)/2} e^{i\v{k}\cdot(\v{d}_1+\v{d}_3)}\nn
&\times t^\perp_{\v{k}+eB\cos\theta' \uv{z}\times (\v{d}_1-\v{d}_3)/4}t^\perp_{\v{k} - eB\cos\theta' \uv{z}\times (\v{d}_1-\v{d}_3)/4}\nn
&\times \tilde{A}^B(\v{d}_1;\epsilon_F)\tilde{A}^B(\v{d}_3;\epsilon_F),
\end{align}
where the phase the particle picks up from the in-plane field has been picked up by the wavevector of the hopping terms. Fourier transforming the $\tilde{A}^B$ terms and interpreting the $\v{d}_1,\v{d}_3$ terms as derivatives in Fourier space allows for the completion of the sums in real space to see that
\begin{align}
\sigma_{zz} \propto& \int \frac{\td^2\v{k}}{(2\pi)^2}\left(t^\perp_{\v{k}}\right)^2 \tilde{A}^B(\v{k}+\v{k}_B/2;\epsilon_F)\tilde{A}^B(\v{k}-\v{k}_B/2;\epsilon_F),
\end{align}
where the $eB\cos\theta$ terms vanish.

\section{Linear response results for a parallel magnetic field}
\label{sec:lin_resp_parallel}

If the magnetic field is parallel to the layers, then there will be no out-of-plane component to induce oscillations in $\tilde{A}^B$. Thus, 
\begin{equation}
A(\v{k},\epsilon_F) = \frac{Z_{\v{k}}\Gamma_{\v{k}}}{(\epsilon_{\v{k}}-\epsilon_F)^2 + \Gamma_{\v{k}}^2},
\end{equation}
where $\Gamma_{\v{k}}$ is the $\v{k}$-dependent scattering rate. Since we assume that $\Gamma/\epsilon_F \ll 1$, Eq.~\ref{eqn:final_Kubo} will become an equation with two sharply peaked Lorentzians. This can be expanded in powers of the magnetic field via the use of a bandwidth expansion. However, it turns out that this will be equivalent to simply reducing one of the spectral functions to a delta function, such that
\ \\
\begin{align}
\sigma_{zz}(90^{\circ},\phi') \propto& \int \frac{\td^2 \v{k}}{(2\pi)^2}  \frac{Z_{\v{k}+\v{k}_B}\Gamma_{\v{k}+\v{k}_B}}{(\epsilon_{\v{k}+\v{k}_B}-\epsilon_F)^2 + \Gamma_{\v{k}+\v{k}_B}^2} \nonumber \\
&\times \left(t^\perp_{\v{k}+\v{k}_B/2}\right)^2 Z_{\v{k}}\delta(\epsilon_{\v{k}}-\epsilon_F).
\end{align}
Since $\Gamma,t^\perp$ and $Z$ have an angular variation around the Fermi surface their magnetic field corrections will be order $k_B/k_F$. In contrast, on the Fermi surface $(\epsilon_{\v{k} + \v{k}_B}-\epsilon_F)/\Gamma_{\v{k}} = \v{v}_F(\phi) \cdot \v{k}_B\tau_\phi + \cdots$, where $\phi$ is the angle along the Fermi surface. This will be of the order $k_Bl$, where $l\sim v_F\tau$ is the in-plane mean free path. Thus the contribution from the correction to the energy will be larger than those to $\Gamma, t^\perp$ and $Z$ by a factor of $k_Fl\gg 1$. Including the necessary factor of the density of states, $|\v{v}_F(\phi)|^{-1}$ this will then become
\begin{equation}
\sigma_{zz}(90^{\circ},\phi') \propto \int \frac{\td \phi}{|\v{v}_F(\phi)|} \frac{\left(t^\perp_\phi Z_\phi\right)^2 \tau_\phi}{1+ \left(\tau_\phi \v{v}_F(\phi) \cdot \v{k}_B\right)^2}.
\end{equation}
We can remove the factor of $Z_\phi$ provided we are conscious that $t^\perp_\phi$ must necessarily be renormalized by it.

\section{ADMR with full three-dimensional $(\pi,\pi)$ reconstruction}

\begin{figure}
\centering
\includegraphics[width=\columnwidth]{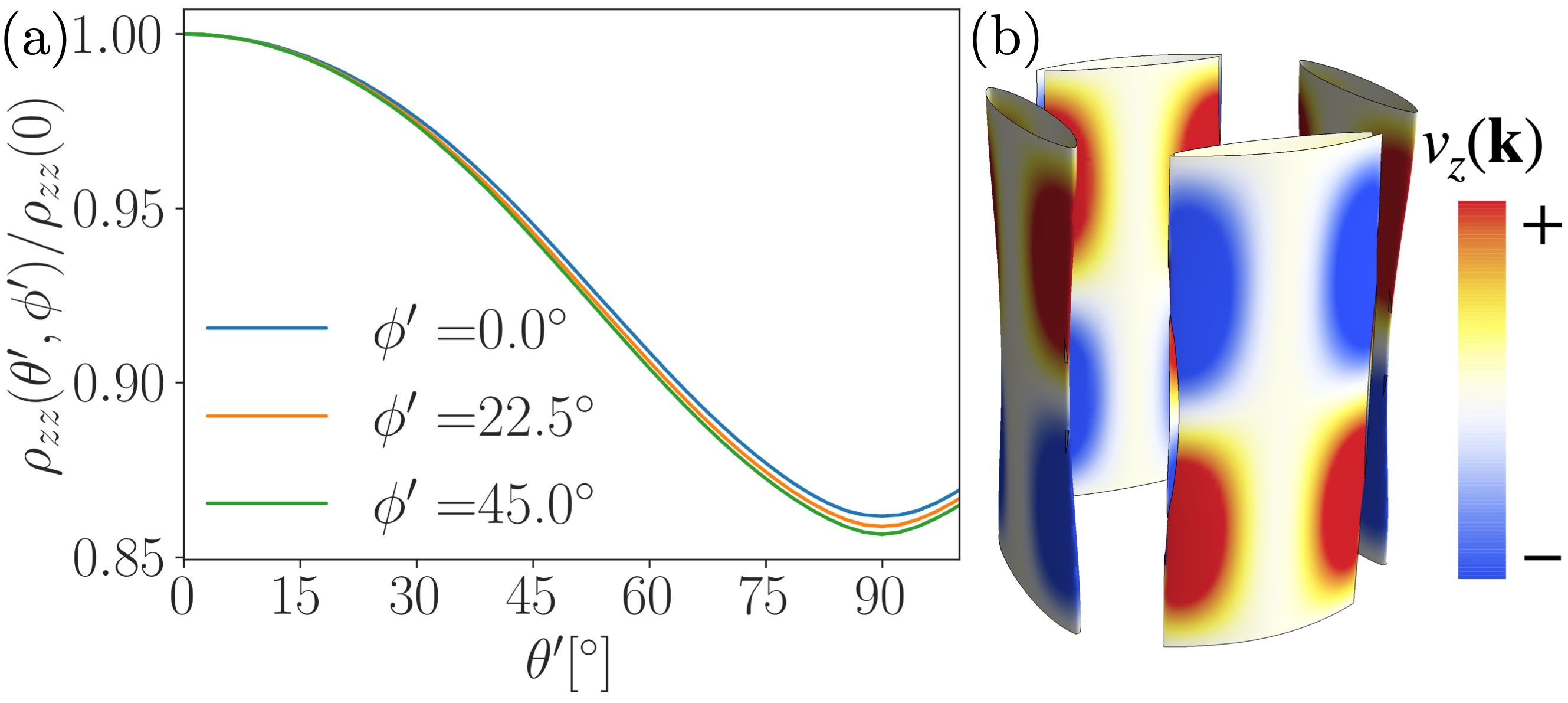}
\caption{ADMR curve for the fully three-dimensional $(\pi,\pi)$ reconstruction (a) The ADMR curve when the three-dimensional dispersion, including the term $t^\perp_{\v{k}}\cos(k_za)$, is reconstructed. Note its similarity to Figure \ref{fig:Zk_inclu}(a) (b) The discretized Fermi surface. The colors represent $v_z(\v{k})$. Note the breaking of $C_4$ symmetry, as here $\langle t^\perp_\phi\rangle$ will change sign around two pockets though not the other two. This is due to the $\cos(k_xa_{\parallel}/2)\cos(k_ya_{\parallel}/2)$ term in $t^\perp_{\v{k}}$.}
\label{fig:3d_pipi}
\end{figure}

The full three dimensional Fermi surface used in \cite{fang_fermi_2020} is defined by setting equation Eq.~\ref{eqn:gen_eps} equal to the Fermi energy, where $\epsilon_{\parallel}(k_x,k_y)$ involves hopping terms on the square lattice up to next-next-nearest neighbor and $t^\perp_{(k_x,k_y)}$ has the form of equation Eq.~\ref{eqn:tperp_k_form}. The $\cos(k_xa_{\parallel}/2)\cos(k_ya_{\parallel}/2)$ term in $t^\perp_{(k_x,k_y)}$ will mean that if a $(\pi,\pi)$ reconstruction is done on the full three-dimensional Fermi surface the $C_4$ rotational symmetry will be broken, as mentioned by \cite{fang_fermi_2020}. Nonetheless, for the same parameters used there we find that the ADMR curve looks similar to the curve when only $\epsilon_{\parallel}(k_x,k_y)$ is reconstructed and $Z_{\v{k}}$ is included, as pictured in Figure \ref{fig:Zk_inclu}(a). The results of the full three-dimensional reconstruction are shown in Figure \ref{fig:3d_pipi}.

Note that we did not take into account the finite correlation length of the ordering along the $c-$direction, since we found it unlikely that a completely uncorrelated ordered SDW along the $c-$direction with a short correlation length would be consistent with the indirect evidence for reasonably coherent semiclassical orbits in ADMR. This is discussed in Section \ref{sec:summary}.

\section{Semiclassical equations for toy model}

For the sake of completeness, we write down the full equations of motion for $\epsilon(k,\phi, k_z)$ for the toy model defined by Eq.~\ref{eqn:toy_model_first}. These are given by $\hbar \dot{\v{k}} = e\v{B}\times \v{v}(\v{k})$, such that
\begin{align}
\dot{k} =& \frac{eB}{\hbar}\left[\sin\theta'\sin(\phi'-\phi) \frac{2a}{\hbar}t^\perp_\phi \sin(k_za)\right. \nn
&\left.+\cos\theta' \frac{2}{\hbar k} \frac{\partial t^\perp_\phi}{\partial \phi} \cos(k_za)\right]\nn
\dot{k}_\phi =& \frac{eB}{\hbar}\left[\cos\theta' \frac{\hbar}{m^*}k\right. \nn
&\left.-\sin\theta'\cos(\phi'-\phi)\frac{2a}{\hbar} t^\perp_\phi \sin(k_za)\right]\nn
\dot{k}_z =& \frac{eB}{\hbar}\left[-\sin\theta'\cos(\phi'-\phi)\frac{2}{\hbar k}\frac{\partial t^\perp_\phi}{\partial \phi} \cos(k_za)\right. \nn
&\left.-\sin\theta'\sin(\phi'-\phi)\frac{\hbar}{m^*}k\right] \label{eqn:full_EOM}.
\end{align}

\section{Role of charge-density wave (CDW) ordering }

In addition to the alternate explanation offered for the $p=0.21$ data in the discussion above, we also investigated a number of possible CDW models. In order to be consistent with the known phenomenology, we  only include the small electron pocket near the nodal region of the large Fermi surface, as was done in \cite{fang_fermi_2020}. We investigated CDWs with period-$2,3,$ and $4$ commensurate order with a constant CDW order parameter, $\vec{\Delta}(\v{k}) = \vec{\Delta}$. We also allowed $\vec{\Delta}$ to be anisotropic, i.e. $\vec{\Delta} = (\Delta_x,\Delta_y)$ with $\Delta_x\neq \Delta_y$ and additionally considered a fully stripe CDW ordering with a period-$3$ commensurate period. For all of these different scenarios we considered both reconstructing the CDW in-plane, with inclusion of $Z_{\v{k}}$, and reconstructing the full three-dimensional Fermi surface. Finally, in order to be consistent with \cite{fang_fermi_2020} we considered CDW reconstructions of this type in-plane with $Z_{\v{k}}=1$. 

In none of these scenarios were we able to reproduce the $p=0.21$ data, even to the level of Figure \ref{fig:v_renorm}(b). The principle challenge for almost all of these CDW orders (and all parameter values tuned in each order) was achieving an upturn of large enough size at $\theta' = 90^{\circ}$. We attribute this to the very small wavevector size $k_F$ of the CDW nodal electron pockets. As our toy model reveals, the size of $\omega_0\tau k_Fa$ is what controls the size of the peak at $\theta' = 90^{\circ}$. Thus the smallness of $k_F$ for the nodal electron pockets meant that in order to see an upturn at $\theta'=90^{\circ}$ the quantity $\omega_0\tau$ had to be made rather large. Since $t^\perp_\phi \geq 0$ for the whole Fermi surface in all of these models, increasing $\omega_0\tau$ always lead to an upturn at $\theta'\ll 1$ before any upturn at $\theta'=90^{\circ}$. Thus the ADMR curves with CDW ordering never resembled the $p=0.21$ curves and for appropriate values of $\omega_0\tau$ instead resembled the $p=0.24$ curves of \cite{fang_fermi_2020}.

\bibliography{ADMR.bib}

\end{document}